\documentclass{aa}  
\usepackage{graphicx}
\usepackage[T1]{fontenc}
\usepackage[utf8]{inputenc}
\usepackage{CJKutf8}
\usepackage[absolute,overlay]{textpos}
\usepackage{subfigure}
\usepackage{natbib}
\usepackage{caption}
\usepackage{rotating}
\usepackage{threeparttable}
\usepackage{titlesec}
\usepackage{placeins}
\usepackage{lscape}
\usepackage{longtable}
\usepackage{needspace}
\usepackage{afterpage}
\usepackage{tabularx}
\usepackage{ltxtable}
\usepackage{float}
\usepackage[switch]{lineno}
\usepackage{amsmath}
\usepackage{appendix}
\usepackage{CJKutf8}
\usepackage[utf8]{inputenc}
\usepackage{xcolor}
\usepackage{ulem}
\usepackage{multirow}

\usepackage{txfonts}

\usepackage{xcolor}

\usepackage{hyperref}
\hypersetup{
    colorlinks=true,    
    linkcolor=blue,     
    citecolor=blue,     
    filecolor=magenta,  
    urlcolor=blue       
}
\usepackage{orcidlink}

\graphicspath{{./}{figures/}}

\begin{document}
\begin{CJK}{UTF8}{gbsn}

\title{AGN Impact on Kiloparsec-scale Star Formation in Nearby Galaxies: Insights from Radio-selected AGNs in ALMaQUEST}

\author{
Pei-Bin Chen \begin{CJK*}{UTF8}{gbsn}(陈培彬)\end{CJK*}\orcidlink{0009-0002-4905-263X}$^{1,2}$\thanks{E-mail: jfwang@xmu.edu.cn},
Xiaoyu Xu \begin{CJK*}{UTF8}{gbsn}(许啸宇)\end{CJK*}\orcidlink{0000-0003-0970-535X}$^{3,4}$,
Junfeng Wang \begin{CJK*}{UTF8}{gbsn}(王俊峰)\end{CJK*}\orcidlink{0000-0003-4874-0369}$^{5}$\thanks{E-mail: chenpb@cwnu.edu.cn},
Taotao Fang \begin{CJK*}{UTF8}{gbsn}(方陶陶)\end{CJK*}\orcidlink{0000-0002-2853-3808}$^{5}$,
Jianfeng Wu \begin{CJK*}{UTF8}{gbsn}(武剑锋)\end{CJK*}\orcidlink{0000-0001-7349-4695}$^{5}$,
Tian-Wen Cao \begin{CJK*}{UTF8}{gbsn}(曹天文)\end{CJK*}\orcidlink{0000-0002-1335-6212}$^{6,7}$,
Mengting Shen \begin{CJK*}{UTF8}{gbsn}(沈梦婷)\end{CJK*}\orcidlink{0009-0003-2072-1200}$^{5}$,
Xiao Cao \begin{CJK*}{UTF8}{gbsn}(曹霄)\end{CJK*}\orcidlink{0000-0002-8411-3314}$^{5}$,
Chunyi Zhang \begin{CJK*}{UTF8}{gbsn}(张春意)\end{CJK*}\orcidlink{0009-0007-7542-1140}$^{5}$, and
Qingshun Hu \begin{CJK*}{UTF8}{gbsn}(胡卿顺)\end{CJK*}\orcidlink{0000-0002-5448-6267}$^{1,2,8}$
\\}
\institute{$^{1}$ School of Physics and Astronomy, China West Normal University, No. 1 Shida Road, Nanchong 637002, People$'$s Republic of China\\
$^{2}$ Local Universe and Time Domain Astronomy Laboratory, China West Normal University, No. 1 Shida Road, Nanchong 637002, People$'$s Republic of China\\
$^{3}$ School of Astronomy and Space Science, Nanjing University, Nanjing 210023, People$'$s Republic of China\\
$^{4}$ Key Laboratory of Modern Astronomy and Astrophysics, Nanjing University, Nanjing 210023, People$'$s Republic of China\\
$^{5}$ Department of Astronomy, Xiamen University, Xiamen, Fujian 361005, People$'$s Republic of China\\
$^{6}$ National Astronomical Observatories, Chinese Academy of Sciences, Beijing 100101, People$'$s Republic of China\\
$^{7}$ Guizhou Radio Astronomical Observatory, Guizhou University, Guiyang 550000, People$'$s Republic of China\\
$^{8}$ Laboratoire d’Astrophysique de Bordeaux, Univ. Bordeaux, CNRS, B18N, all\'{e}e Geoffroy Saint-Hilaire, 33615 Pessac, France
}

\date{Accepted XXX. Received YYY; in original form ZZZ}

\abstract
{Understanding how active galactic nuclei (AGN) regulate star formation (SF) in their host galaxies is central to galaxy formation and evolution. In particular, whether AGN feedback enhances, suppresses, or redistributes SF on (sub)kiloparsec scales remains an open question.}
{We aim to investigate the impact of AGN activity on the molecular interstellar medium (ISM) and star formation efficiency (SFE) by comparing the spatially resolved scaling relations between radio-selected AGN host galaxies and  optically selected star-forming galaxies (SFGs) from ALMaQUEST.}
{We compare the stellar populations (SPs) ages - $D_{\rm n}$4000, the CAS (concentration, asymmetry, smoothness) parameters  of CO(1-0) maps, and the scaling relations of the AGN hosts - including the spatially resolved Kennicutt-Schmidt law (rKS) and the molecular gas main sequence (rMGMS) - with those of star-forming galaxies (SFGs).}
{We found that AGN hosts exhibit younger SPs at 0.4 - 0.6 R$_{\rm e}$ (1 - 2 kpc), and there is no statistical difference between their CAS and SFGs. Although pixels of both AGN hosts and SFGs occupy similar regions on the spatially rMGMS, indicating a gradual decline of gas content with center distance, Seyfert pixels in AGN hosts lie substantially below the star-forming/composite pixels and SFGs on the spatially rKS plane, reflecting a reduced SFE.}
{Analyzing the $D_{\rm n}$4000 and H$\alpha$ suggests that the younger SPs in AGN hosts are caused by the internal structures of galaxies -- spiral arms. The results of the CAS might be the outcome of the interaction of multiple factors including bars, AGN, and resolution. Overall, the near-identical rMGMS combined with the lower SFE of Seyfert pixels within AGN hosts provides evidence for negative AGN feedback.}


\titlerunning{AGN Impact on Kiloparsec-scale Star Formation in Nearby Galaxies}
\authorrunning{Chen et al.}

\maketitle

\section{Introduction} \label{intro} 

Numerous connections exist between the properties of supermassive black holes (SMBHs) and their host galaxies (e.g., the M$_{\rm BH}$ - $\sigma_{\rm bulge}$ relation), indicating the coevolution between their growth and their host galaxies \citep[e.g.,][]{1998AJ....115.2285M, 2000ApJ...539L...9F, 2013ARA&A..51..511K, 2023MNRAS.518.2088M}. BHs grow by accreting cold gas onto their accretion disks. Once this process occurs, they become active, and galaxies are called active galactic nuclei \citep[AGN;][]{2018MNRAS.478.5491M, 2023MNRAS.518.2088M, 2025A&A...694A.309J}. This process will inject a large amount of energy into the host galaxy, which will have an impact on its evolution, known as AGN feedback \citep[e.g.,][]{2012ARA&A..50..455F, 2023ApJ...953...26L}. AGN feedback affects the interstellar medium (ISM) of their hosts through radiation and outflows, heating or ejecting gas, or preventing gas accretion, ultimately suppressing star formation (SF) and the growth of BHs \citep[][]{2007ApJ...671.1388D, 2008ApJS..175..356H, 2013ApJ...772..112S, 2013MNRAS.433.3079Z, 2023ApJ...953...26L, 2024ApJ...971..165X, 2025ApJ...979...26S}.

The AGN feedback is classified into two categories: radiative mode and kinetic mode \citep[e.g.,][]{2012ARA&A..50..455F}. Observations of massive galaxies in high-density environments (galaxy groups or clusters) have revealed that the cooling flows generated by radiation cooling are suppressed, and thus no significant accumulation of stars occurs at the centers of galaxy clusters \citep[][]{2014ARA&A..52..589H, 2025ApJ...979...26S}. This indicates that there might be certain mechanisms at the center that can heat the circumgalactic medium of the galaxy cluster, thereby suppressing the cooling flows. The most likely heating source is the AGN in massive galaxies at their centers, which are in a state of lower accretion rate (lower Eddington ratio, $L_{\rm bol}/L_{\rm Edd}$\footnote{The ratio of the AGN bolometric luminosity ($L_{\rm bol}$) to its Eddington luminosity ($L_{\rm Edd}$), see Sect. \ref{3.4}.}) and mainly release energy through jets \citep[i.e. kinetic mode;][]{2014ARA&A..52..589H, 2014ARA&A..52..529Y, 2018MNRAS.479.5544M, 2022NatAs...6..109M, 2025ApJ...993...35X}. However, when it has a relatively high $L_{\rm bol}/L_{\rm Edd}$ ($\geq$ 0.01), the main way to release energy is through electromagnetic radiation, which theoretically can drive a strong outflow of gas (molecular and ionized gas). This pattern is called radiative mode \citep[e.g.,][]{2012ARA&A..50..455F, 2015ARA&A..53..115K, 2015Natur.519..436T, 2024A&A...681A..63S, 2024ApJ...971..165X}.

AGNs can heat or remove cold gas or prevent gas accretion, resulting in the depletion of the raw material for SF and thereby suppressing SF activities. This is known as AGN negative feedback \citep[e.g.,][]{2006ApJS..163....1H, 2008ApJS..175..356H, 2012RAA....12..917S, 2019MNRAS.486.2827D}. \citet{2023ApJ...953...26L} found that the low-redshift AGN hosts have significantly suppressed central star formation rates (SFRs) on kiloparsec (kpc) scales, with 2 times lower SFR than their control sample. \citet{2025A&A...694A.309J} suggested that the observed present-day radio AGN activity is not responsible for the past quenching of their hosts but may help their hosts maintain quiescence by ionizing and heating the surrounding gas. However, some studies have also suggested that during gas-rich periods, AGN outflows can excessively compress the gas, thus enhancing SF in the corresponding regions. This is known as AGN positive feedback \citep[e.g.,][]{2007ApJ...671.1388D, 2013ApJ...772..112S, 2023A&A...678A.127V, 2024ApJ...971..165X}. \citet{2007ApJ...671.1388D} found evidence of a young stellar population (SP, 10 - 300 Myr) in the central regions of several nearby Seyfert galaxies. Recently, \citet{2024ApJ...971..165X} investigated the Seyfert galaxy NGC 2992, an AGN known for its significant variability, suggesting that there may be two effects (negative and positive) simultaneously in the circumnuclear regions.

The SF of galaxies involves extremely complex processes that convert molecular gas into stars, including gas accretion, collapse, and cooling of molecular clouds \citep[][]{2014ApJ...796..135K}. The conversion of gas to stars can be well expressed using the Kennicutt-Schmidt law \citep[KS law;][]{1959ApJ...129..243S, 1998ApJ...498..541K, 2012ARA&A..50..531K}, which establishes a direct relationship between the SFR surface density ($\Sigma_{\rm SFR}$) of galaxies and their gas content (molecular gas mass surface density, $\Sigma_{\rm mol}$) in the form of a power law equation ($\Sigma_{\rm SFR}$ $\varpropto$ $\Sigma_{\rm mol}^{\rm \alpha}$). This relationship can help us to understand the feedback from AGN \citep[e.g.,][]{2007ApJ...671..333K, 2015A&A...577A.135C, 2021ApJ...913..139G, 2023ApJ...944...30M, 2024ApJ...971..165X}. Although molecular hydrogen (${\rm H_{2}}$) serves as a raw material for SF, its unique structure makes direct observation challenging \citep[e.g.,][]{1991ARA&A..29..581Y, 2005ARA&A..43..677S, 2025ApJ...978...59C}. Numerous studies have shown that carbon monoxide (CO) can be effectively traced ${\rm H_{2}}$, and the conversion factor ($\alpha_{\rm CO}$) is used to directly link the luminosity of CO with the mass of ${\rm H_{2}}$ \citep[e.g.,][]{1991ARA&A..29..581Y, 2003ApJ...588..771Y, 2005ARA&A..43..677S, 2013ARA&A..51..207B}. 

The index ($\alpha$) of the KS law ranges from 0.5 to 3 \citep[][]{2008AJ....136.2846B}, depending not only on the SFR and the gas tracer \citep[][]{2004ApJ...606..271G}, but also on the physical scales used to calculate the surface density \citep[][]{2018ApJ...863L..21K}. For example, on a large scale (the entire galaxy), $\alpha$ is super-linear \citep[][]{2004ApJ...606..271G}, while on a small scale (kpc or sub-kpc) it approaches 1 or even lower \citep[][]{2008AJ....136.2846B, 2019ApJ...884L..33L, 2024MNRAS.52710201E}. Advancements in relevant instrumentation have led to significant research focusing on the spatially resolved KS law (hereafter, rKS), which measures SFR and gas properties at kpc/sub-kpc scales \citep[e.g.,][]{2007ApJ...671..333K, 2008AJ....136.2846B, 2019ApJ...884L..33L, 2021ApJ...913..139G, 2023ApJ...944...30M, 2024MNRAS.52710201E, 2025PASA...42..140S}. \citet{2007ApJ...671..333K} found $\alpha$ for M51a on the 0.5-2.0 kpc scale to be about 1.37 $\pm$ 0.03, and then the discrepancy between the linear and super-linear was explained by \mbox{\citet{2011ApJ...735...63L}}. \citet{2015A&A...577A.135C} studied 4 low-luminosity AGNs (LLAGNs) and found that each galaxy is characterized by a distinct molecular SF relation on spatial scales between 20-200 pc. \citet{2021ApJ...913..139G} combined Atacama Large Millimeter Array (ALMA) and Multi Unit Spectroscopic Explorer (MUSE) data to investigate the molecular and ionized gas in the central 5.4 kpc region of NGC 1365, and found that the SFR in the inner circumnuclear ring is about 0.4 dex higher than that in the outer regions, and at a linear resolution of 180 pc, the super-linear rKS ($\alpha$ $\sim$ 1.96 $\pm$ 0.14) are obtained ($\Sigma_{\rm mol}$ $>$ 1 $M_{\odot}{\rm pc}^{-2}$). Recently, \citet{2023ApJ...944...30M} studied ongoing SF in 6 low-redshift Paloma-green quasar (PG Quasar) at kpc scale, and found that the rise of $\Sigma_{\rm SFR}$ toward the nucleus is steeper than that of $\Sigma_{\rm mol}$ (up to 0.15-0.80 $M_{\odot}$yr$^{-1}$kpc$^{-2}$), and that the gas in the nuclear regions is converted to stars with a shortened depletion time (0.2-2.0 Gyr), indicating that these regions can be deemed as starbursts (see their figure 6). The study of NGC 2992 by \citet{2023A&A...679A..88Z} suggested that its rKS is consistent with the local AGN hosts \citep[][]{2015A&A...577A.135C}.

However, we note that the rKS in these studies is related either to the distribution of a specific region within a galaxy at a given scale (global or kpc/sub-kpc) or to the distributions of all the pixels across the galaxy. Using classical Baldwin, Phillips $\&$ Telervich \citep[BPT;][]{1981PASP...93....5B} diagnosis\footnote{The theoretical boundaries of the BPT diagram come from \cite{2003MNRAS.346.1055K}, \cite{2001ApJ...556..121K}, and \cite{2007MNRAS.382.1415S}, respectively.\label{curve}} ([O\,{\sc iii}]$\lambda$5007/H$\beta$ vs [N\,{\sc ii}]$\lambda$6583/H$\alpha$), it is possible to classify galaxies (Star-forming, Composite or Seyfert). With the advent of integral field unit (IFU) data (e.g., Mapping Nearby Galaxies of the Sloan Digital Sky Survey at Apache Point Observatory, SDSS-IV MaNGA), this classification also applies to individual pixels within the galaxy. The pixel classification allows us to focus only on a specific type of pixel or to compare the effects of AGN on three types of pixels within their hosts.  In this work, we select galaxies from the ALMA-MaNGA QUEnching and STar formation \citep[ALMaQUEST\footnote{\url{http://sites.google.com/view/almaquest/home}};][]{2020ApJ...903..145L} survey and compare AGN hosts with star-forming galaxies (SFGs) in terms of SPs, non-parametric morphological parameters of CO(1-0) maps, and scaling relations (e.g., rKS). We aim to investigate the effects of AGN on both the entire galaxy and the different types of pixels within galaxies. 

The organizational structure of this paper is as follows. In Sect. \ref{data}, we briefly introduce the data and sample selection. The calculation methods for the relevant quantities are provided in Sect. \ref{Methodology}. Our results and discussion are provided in Sects. \ref{result} and \ref{discussion}. In Sect. \ref{conclusions}, we provided relevant conclusions. Throughout this paper, we adopt a set of cosmological parameters as follows: $H_0$ = 70 km s$^{-1}$ Mpc$^{-1}$ ($h$ = 0.7), $\Omega_{\rm m}$ = 0.30, $\Omega_{\Lambda}$ = 0.70.

\section{Data and Sample Selection} 
\label{data}

\begin{figure*}[!ht]
    \centering
    \includegraphics[width=\linewidth]{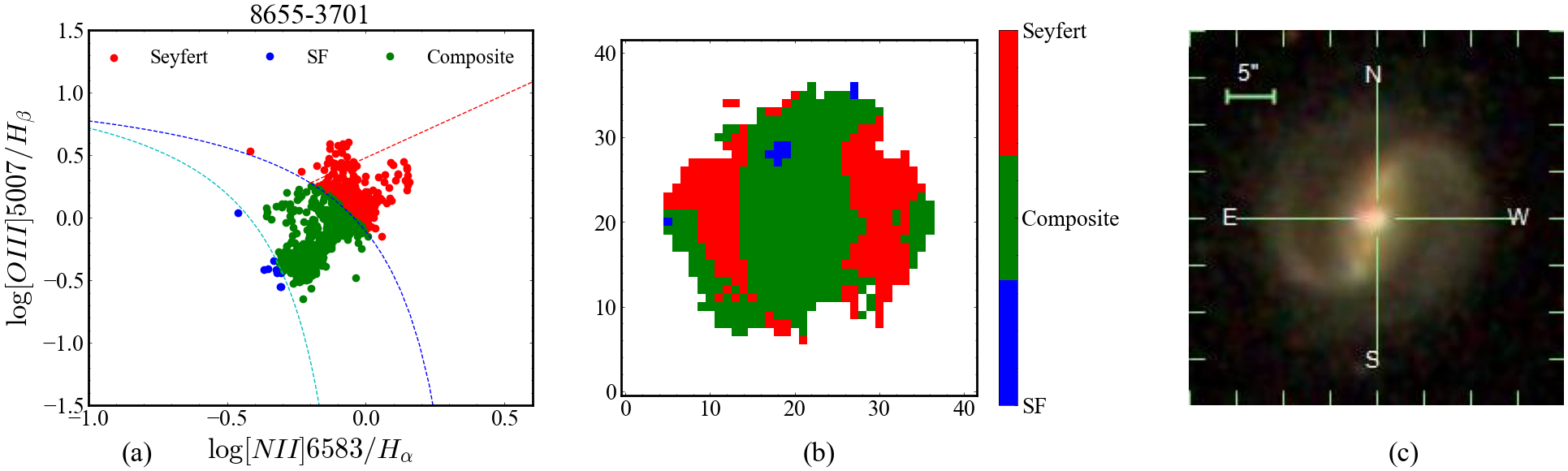}
    \caption{Example of spatially resolved BPT diagnosis for AGN host galaxy 8655-3701. a) Distributions of each pixel in the BPT diagram. The cyan, blue and red dotted theoretical lines correspond to \cite{2003MNRAS.346.1055K}, \cite{2001ApJ...556..121K} and \citet{2007MNRAS.382.1415S} respectively. b) The spatial distribution of each pixel. c) SDSS g/r/i color image of galaxy 8655-3701. In this Figure, blue, green, and red dots represent pixels in the star-forming, composite, and Seyfert regions, respectively.}
    \label{spatially BPT}
\end{figure*}

\subsection{Data from ALMaQUEST}\label{sec2.1}

ALMaQUEST consists of 46 galaxies selected from the SDSS-IV MaNGA \citep[][]{2015ApJ...798....7B, 2016AJ....152..197Y} with complementary observations made by the ALMA of CO(1-0). ALMaQUEST compiles data sets from four individual ALMA programs - 2015.1.01225.S, 2017.0.01093.S, 2018.1.00558.S (PI: L. Lin), and 2018.1.00541.S (PI: Ellison), which aims to investigate the dependence of SF activities on the cold molecular gas content on kpc scales in nearby galaxies \citep[][]{2020ApJ...903..145L}. These galaxies have a wide range of specific SFR (sSFR), spanning from the main sequence (MS), the green valley, and up to the starburst regimes. These galaxies are mostly in the redshift interval 0.02 $<$ z $<$ 0.05, over which range 1$^{\prime\prime}$ corresponds to $\sim$ 0.5 - 1.0 kpc. ALMA observations were designed to deliver a primary beam size (angular resolution) that is well-matched to the MaNGA point spread function ($\sim$ 2.5$^{\prime\prime}$), allowing the ALMA data cubes to be projected onto the same spatial grids \citep[][]{2019ApJ...884L..33L, 2020ApJ...903..145L, 2021MNRAS.501.4777E, 2022ApJ...926..175L}. The properties of the sample, the reduction of data and the products are fully described in the main survey paper \citep[][]{2020ApJ...903..145L}. Our CO(1-0) data come from the ALMaQUEST survey and require the signal-to-noise ratio (S/N) $>$ 3$\sigma$. 
The information of the remaining optical emission lines (e.g., H$\alpha$, [O\,{\sc iii}]$\lambda$5007, [N\,{\sc ii}]$\lambda$6583, H$\beta$) come from the Pipe3D \citep[][]{2016RMxAA..52..171S, 2016RMxAA..52...21S, 2022ApJS..262...36S} data cubes \footnote{\url{https://data.sdss.org/sas/dr17/manga/spectro/pipe3d/v3_1_1/3.1.1/}}, and they are required to have S/N $>$ 3. Our global stellar mass comes from GALEX-SDSS-WISE Legacy Catalog \citep[GSWLC-A2 \footnote{\url{https://salims.pages.iu.edu/gswlc/}};][]{2016ApJS..227....2S, 2018ApJ...859...11S}, while stellar mass surface density ($\Sigma_{\ast}$) from Pipe3D data cubes. Furthermore, we extract the parameter $D_{\rm n}$4000 from the same cube as a proxy for the age of the galaxy stellar population. It is the average flux density ratio of the bands 4000-4100 \,\AA\ and 3850-3950 \,\AA.
The other parameters (e.g., axis ratio, R$_{\rm e}$, redshift) come from the NASA-Sloan Atlas (NSA) catalog \citep[v1$\_$0$\_$1 \footnote{\url{https://nsatlas.org/data}\label{NSA}};][]{2011AJ....142...31B}. 

\subsection{Galaxies from ALMaQUEST}\label{sec2.2}

The accurate identifications of AGNs and the spatially resolved characteristics of host galaxies are powerful combinations for studying the role of AGNs in the co-evolution of galaxies and their central SMBHs \citep[e.g.,][]{2020ApJ...901..159C, 2024ApJ...963...53C}. \cite{2020ApJ...901..159C} presented robust identifications of 406 AGNs in the first 6261 galaxies observed by the MaNGA survey, using multi-band data (WISE mid-infrared colors, Swift/BAT ultra-hard X-ray, NVSS and FIRST radio, and broad emission lines in SDSS spectra) instead of the usual optical line flux ratios to exclude the influence of shocks and metallicity. This catalog is provided as a MaNGA value-added catalog (hereafter MaNGA$\_$AGN\footnote{\url{https://www.sdss4.org/dr17/data_access/value-added-catalogs/?vac_id=the-manga-agn-catalog}}) and is described in detail in \cite{2020ApJ...901..159C} and \cite{2024ApJ...963...53C}. We cross-reference the ALMaQUEST catalog with the MaNGA$\_$AGN to obtain 6 AGN hosts, which are listed in Table \ref{tab1} (marked with stars). According to the MaNGA$\_$AGN, we find that these 6 AGN hosts are all radio-selected AGN.

The radio luminosities of our AGN hosts (Table \ref{tab1}) are greater than 43.5 erg/s \citep[log $L_{\rm radio}$, NVSS;][]{2020ApJ...901..159C}, and they do not exhibit radio jet characteristics under the resolution of VLA NVSS or FIRST survey. In addition, according to \citet{2012MNRAS.421.1569B}, they are all classified as radio-quiet galaxies. Therefore, we suggest that their pixels classified as Seyfert (Fig. \ref{spatially BPT}) in these 6 AGN hosts represent the remnants of previous AGN activities \citep[e.g.,][]{2025A&A...694A.309J}. We perform a spatially resolved BPT diagnosis$^{\ref{curve}}$ in each galaxy (Table \ref{tab1}) to distinguish the type of each pixel (Star-forming, Composite or Seyfert). Fig. \ref{spatially BPT} is an example of one of our AGN host galaxy 8655-3701, and the optical image represents the SDSS g, r, i color image. In Appendix \ref{remaining BPT}, we will present its spatially resolved BPT diagrams for the remaining 5 AGN hosts. However, we found that only a few pixels in the three AGN hosts fell within the Seyfert regime (one in 8083-12702; two in 8615-3703; nine in 8616-9102), and these pixels are at the edges of the data far from the true location of the possible AGN at the galaxy center (therefore more likely to be extended emission line regions or poor quality data, not true Seyferts). According to \citet{2020ApJ...901..159C}, we notice that these AGN hosts are identified by radio luminosity \citep[NVSS and FIRST;][]{2012MNRAS.421.1569B}, so they may not necessarily be active in the optical band.

We have noticed that pixels classified as Seyfert are located in the outer regions of the galaxy, especially in galaxies 8655-3701 and 8081-9101. According to Pipe3D, we found that the EW$\_$H$\alpha$ of these 6 AGN hosts are all greater than 3\AA\ \citep{2010MNRAS.403.1036C, 2020ARA&A..58...99S}. Therefore, it is possible to largely rule out the pollution caused by old SPs. For comparison, we select SFGs from the remaining 40 galaxies in the ALMaQUEST survey. We performed a global BPT diagnosis ([O\,{\sc iii}]$\lambda$5007/H$\beta$ vs [N\,{\sc ii}]$\lambda$6583/H$\alpha$ $^{\ref{curve}}$) on the remaining 40 galaxies by integrating the emission line flux within an effective radius (extract from Pipe3D cubes) of each galaxy and require an equivalent width of H$\alpha$ (EW$\_$H$\alpha$ $>$ 6\AA\ ) within an effective radius \citep[e.g.,][]{2020ARA&A..58...99S, 2024A&A...691A.107C, 2025ApJ...978...59C}. We obtain 17 SFGs, listed in Table \ref{tab1}.

\begin{table*}
\caption{Basic information of our galaxies from ALMaQUEST\label{tab1}}
\small
\centering 
\begin{tabular}{cccccccccc} 
\hline\hline 
MANGA$\_$ID & R.A. & Decl. & Redshift & R$_{\rm e}$ & log(M$_{\ast}/M_{\odot})$ & logsSFR & $S_{\rm CO(1-0)}$ & logM$_(\rm H_{2})$ & log$L_{\rm bol}/L_{\rm Edd}$\\
(Plate-IFU) & (deg.) & (deg.) & (z) & (arcsec) & (M$_{\odot}$) & (yr$^{-1}$) & (Jy km s$^{-1}$) & (M$_{\odot}$) & (error)\\
(1) & (2) & (3) & (4) & (5) & (6) & (7) & (8) & (9) & (10)\\
\hline 
8078-6103$^{*}$ & 42.416542 & $-0.069851$ & 0.029 & 6.42 & 10.80 & $-10.29$ & 23.16 $\pm$ 0.12 & 9.558 $\pm$ 0.002 & $-6.02 \pm 0.71$\\
8081-9101$^{*}$ & 47.772182 & $-0.546538$ & 0.028 & 6.92 & 10.73 & $-10.32$ & 18.89 $\pm$ 0.15 & 9.469 $\pm$ 0.003 &  $-6.67 \pm 0.65$\\
8083-12702$^{*}$ & 50.245415 & $-0.367683$ & 0.021 & 28.86 & 11.24 & $-10.73$ & 34.22 $\pm$ 0.14 & 9.468 $\pm$ 0.002 &  $-6.83 \pm 0.59$\\
8615-3703$^{*}$ & 320.82642 & 1.25498 & 0.018 & 4.21 & 10.23 & $-9.81$ & 23.18 $\pm$ 0.22 & 9.175 $\pm$ 0.004 &  $-5.78 \pm 0.60$\\
8616-9102$^{*}$ & 322.74945 & 0.000594 & 0.030 & 7.70 & 10.44 & $-9.81$ & 15.68 $\pm$ 0.07 & 9.446 $\pm$ 0.002 &  $-6.19 \pm 0.62$\\
8655-3701$^{*}$ & 356.75183 & $-0.447387$ & 0.071 & 4.05 & 11.27 & $-10.11$ & 25.68 $\pm$ 0.27 & 10.416 $\pm$ 0.005 &  $-6.23 \pm 0.53$\\
\hline 
7977-3703 & 333.052032 & 12.205191 & 0.028 & 4.72 & 10.43 & $-10.04$ & 8.86 $\pm$ 0.04 & 9.119 $\pm$ 0.002 &  -\\
8077-6104 & 42.032784 & $-0.752316$ & 0.046 & 6.73 & 10.75 & $-10.14$ & 8.07 $\pm$ 0.03 & 9.524 $\pm$ 0.002 &  -\\
8081-3704 & 49.821442 & $-0.969631$ & 0.054 & 3.11 & 10.33 & $-9.42$ & 2.73 $\pm$ 0.01 & 9.195 $\pm$ 0.002 &  -\\
8082-6103 & 49.782173 & 0.955959 & 0.024 & 5.74 & 10.20 & $-9.85$ & 7.47 $\pm$ 0.02 & 8.920 $\pm$ 0.001 &   -\\
8082-12701 & 48.896458 & $-1.016286$ & 0.027 & 9.84 & 10.53 & $-10.46$ & 6.00 $\pm$ 0.02 & 8.925 $\pm$ 0.001 & - \\
8084-3702 & 50.636642 & $-0.001213$ & 0.022 & 5.12 & 10.43 & $-9.94$ & 21.34 $\pm$ 0.13 & 9.294 $\pm$ 0.003 & - \\
8155-6102 & 52.621368 & 0.752068 & 0.031 & 6.66 & 10.38 & $-10.05$ & 9.55 $\pm$ 0.03 & 9.243 $\pm$ 0.001 & -\\
8156-3701 & 55.592297 & $-0.583196$ & 0.052 & 4.12 & 10.27 & $-9.46$ & 1.58 $\pm$ 0.01 & 8.934 $\pm$ 0.002 & - \\
8241-3703 & 126.461189 & 18.166689 & 0.029 & 5.28 & 10.23 & $-9.99$ & 4.14 $\pm$ 0.02 & 8.830 $\pm$ 0.002 & - \\
8241-3704 & 126.568909 & 17.362452 & 0.066 & 5.03 & 11.06 & $-9.84$ & 12.62 $\pm$ 0.09 & 10.039 $\pm$ 0.003 & - \\
8450-6102 & 171.748840 & 21.141676 & 0.042 & 5.75 & 10.31 & $-9.68$ & 4.75 $\pm$ 0.02 & 9.213 $\pm$ 0.002 &  -\\
8615-12702 & 320.159454 & 1.047277 & 0.021 & 19.13 & 10.21 & $-10.10$ & 5.27 $\pm$ 0.02 & 8.645 $\pm$ 0.001 & - \\
8616-6104 & 322.980530 & 0.213767 & 0.054 & 6.63 & 10.90 & $-10.69$ & 4.26 $\pm$ 0.02 & 9.392 $\pm$ 0.002 & - \\
8618-9102 & 319.271454 & 9.972303 & 0.043 & 4.89 & 10.47 & $-10.05$ & 5.52 $\pm$ 0.04 & 9.306 $\pm$ 0.003 & - \\
8623-6104 & 311.780975 & 0.300461 & 0.096 & 3.19 & 11.34 & $-10.14$ & 6.32 $\pm$ 0.04 & 10.077 $\pm$ 0.003 & - \\
8655-9102 & 358.221924 & $-0.382447$ & 0.045 & 4.22 & 10.31 & $-10.06$ & 1.93 $\pm$ 0.01 & 8.884 $\pm$ 0.002 & - \\
8952-6104 & 204.933975 & 27.776474 & 0.028 & 7.39 & 10.60 & $-10.15$ & 8.28 $\pm$ 0.03 & 9.111 $\pm$ 0.001 &  -\\
\hline
\end{tabular}
\\
Notes: In this table, the galaxies marked with stars are our AGN hosts, while the rest are SFGs. The columns (4) and (5) come from the SDSS-IV MaNGA Survey, column (6) from \cite{2018ApJ...859...11S}, and the columns (7) - (9) come from \cite{2021MNRAS.501.4777E} and \cite{2020ApJ...903..145L}. The 10th column represents their log$L_{\rm bol}/L_{\rm Edd}$, and corresponding errors (see Sect. \ref{3.4}). For SFGs, the central BHs are weak or inactive, and the luminosity is not dominated by the BHs. Therefore, we do not calculate them.
\end{table*}

\section{Methodology}\label{Methodology}

\subsection{Extinction Correction}\label{extinction}

To obtain the derived quantities (e.g., SFR) of each pixel, we need to consider dust extinction, which has been exploited in many studies (e.g., \citealt{2000ApJ...533..682C, 2001PASP..113.1449C, 2024A&A...691A.107C, 2025ApJ...978...59C}). \cite{2001PASP..113.1449C} estimated the extinction-corrected H$\alpha$ luminosity using the H$\alpha$/H$\beta$ ratio:

\begin{equation}
    f_{\rm i}(\lambda) = f_{\rm o}(\lambda) \times 10^{\rm 0.4E(B-V)_{\rm gas}\kappa^{\rm e}(\lambda)}
\end{equation}
where $f_{\rm i}(\lambda)$ is the intrinsic flux density, $f_{\rm o}(\lambda)$ is the observed flux density, and $\kappa^{\rm e}(\lambda)$ is the Galactic dust attenuation curve \citep{2000ApJ...533..682C}. The color excess E(B $-$ V) is:

\begin{equation}
    E(B-V) = 2.15 \times {\rm log[(L_{H{\alpha}}/L_{H{\beta}})/(L_{H{\alpha}}/L_{H{\beta}})_{\rm i}]}
\end{equation}
where $L_{\rm H{\alpha}}/L_{\rm H{\beta}}$ is the observed ratio, $(L_{\rm H{\alpha}}/L_{\rm H{\beta}})_{\rm i}$ is an intrinsic Balmer decrement. For the classic H II regions, $(L_{\rm H{\alpha}}/L_{\rm H{\beta}})_{\rm i}$ = 2.86 corresponds to Case B recombination with a temperature of $10^4$ K and electron density $n_{\rm e}$ = $10^2$ cm$^{-3}$ \citep{1989S&T....78..491O}. . However, in AGNs, there are usually higher ionization parameters, higher electron temperatures, electron densities, and collisional excitation. Therefore, here we adopt a more appropriate $(L_{\rm H{\alpha}}/L_{\rm H{\beta}})_{\rm i}$ = 3.10 \citep[][]{2006agna.book.....O}.

Therefore, the SFR of each pixel identified as the star-forming regions by the BPT diagram can be estimated from the extinction-corrected H$\alpha$ luminosity as presented by \cite{1998ARA&A..36..189K}:
\begin{equation}\label{SFR cal}
  {\rm SFR} (\rm M_{\odot}yr^{-1}) = 7.9 \times 10^{-42} L_{\rm H{\alpha}}
\end{equation}

Numerous studies have shown that AGN feedback can suppress or enhance SF in their hosts \citep[e.g.,][]{2008ApJS..175..356H, 2012RAA....12..917S, 2023RAA....23a5005C, 2024ApJ...971..165X}. Using the formula \ref{SFR cal} to calculate the SFR of pixels located in the Composite or Seyfert regions will lead to overestimation. To quantify the contribution of AGN photoionization to the observed H$\alpha$ emission, we employ the re-projected optical diagnostic diagram developed by \citet{2020MNRAS.499.5749J}. \citet{2020MNRAS.499.5749J} presented a new 3D diagnostic diagram, which can be applied to estimate the contributions from AGN and SF based on a given model. They used the best-fitting SF and AGN model for their MaNGA spaxel sample and derived a relation between AGN's contribution and the factor P$_{1}$, which can be approximated as follows:

\begin{equation}\label{for4}
\text{f$_{\rm AGN}$}=
\begin{cases}
    0,\quad P_{1} \leq -0.53\\
    0.14P_{1}^{2} + 0.96P_{1} + 0.47,\quad -0.53 < P_{1} < 0.51\\
    1,\quad P_{1} \geq 0.51
\end{cases}
\end{equation}
where f$_{\rm AGN}$ is the contribution of AGN to the H${\alpha}$ flux (H${\alpha}$$_{\rm AGN}$/H${\alpha}$$_{\rm total}$) and P$_{1}$ is the ratio of emission lines within each pixel (0.63 log([NII]/H${\alpha}$) + 0.51 log([SII]/H${\alpha}$) + 0.59 log([OIII]/H${\beta}$)). By calculating P$_{1}$ and using formula \ref{for4}, we can decompose the observed H${\alpha}$, and consequently obtain AGN-corrected H$\alpha$ luminosities and SFRs for pixels located in Composite and Seyfert regions \citep[e.g.,][]{2021ApJ...923....6J, 2025A&A...694A.309J}. In Sect. \ref{discussion}, we will discuss the impact of this method on our results.

\begin{figure*}
\centering
\includegraphics[width = 0.65\linewidth]{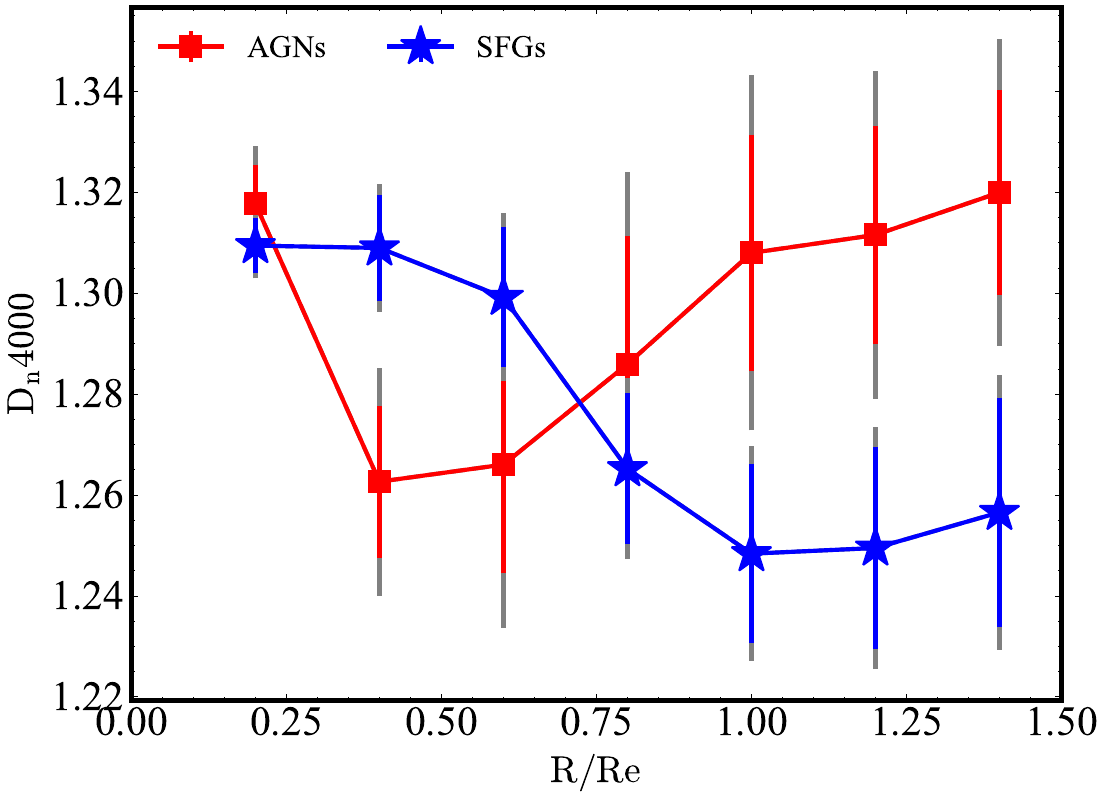}
\caption{Radial profiles of $D_{\rm n}4000$. In this figure, the red and blue lines represent the radial profiles for AGN hosts and SFGs, respectively. In this figure, the colored error bars show the standard deviation of the median of each galaxy within the corresponding radius bin, while the gray error bars show the standard deviation of the sample median profiles across Monte Carlo realizations, in which individual $D_{\rm n}4000$ measurements were perturbed according to their observational uncertainties and both the individual-galaxy and sample medians were recomputed.}
\label{total_dn4000}
\end{figure*}

\subsection{Molecular Gas Mass}\label{H2 mass}

The ALMA data cubes provide the emission line intensity ($I$) in Jy beam$^{-1}$ units. After considering the area of the pixel and the solid angle of the beam \citep[e.g.,][]{2024ApJ...967L..25Z}, we can obtain the converted intensity ($I^{'}$, Jy pixel$^{-1}$):

\begin{equation}
    I^{'}\ (\rm Jy\ pixel^{-1}) = \frac{I\ (\rm Jy\ beam^{-1}) \times (\rm pixel \times pixel)}{1.133 \times \Theta_{\rm x} \times \Theta_{\rm y}}
\end{equation}
where the pixel $\times$ pixel is the pixel area (arcsec$^{2}$); $\Theta_{\rm x}$ and $\Theta_{\rm y}$ are the Beam major and Beam minor, respectively. We integrate the line intensity ($I^{'}$) within the observed velocity range to obtain the integrated line flux ($S_{\rm CO}$, Jy km s$^{-1}$), so the CO(1-0) luminosity can be calculated as \citep[e.g.,][]{1997ApJ...478..144S, 2017AJ....154..116C}:

\begin{equation}
    L_{\rm CO(1-0)} = 3.25 \times 10^{7} \times \frac{S_{\rm CO(1-0)}D_{\rm L}^{2}}{\nu_{\rm obs}^{2}(1 + z)^{3}}
\end{equation}
where $\nu_{\rm obs}$ is the observed frequency of the CO(1-0) line in GHz; $D_{\rm L}$ is the luminosity distance in units of Mpc; and the unit of $L_{\rm CO(1-0)}$ is K km s$^{-1}$ pc$^{2}$. 

The molecular hydrogen mass ($M_{\rm H_2}$) of each pixel can be calculated as \citep[e.g.,][]{2011MNRAS.415...32S, 2017AJ....154..116C}:
\begin{equation}
    M_{\rm H_2} = L_{\rm CO(1-0)} \times \alpha_{\rm CO(1-0)}
\end{equation}
where $\alpha_{\rm CO(1-0)}$ is the mass-to-light ratio that depends on the metallicity. We adopt Equation 4 of \cite{2020ApJ...892..148S} to calculate $\alpha_{\rm CO(1-0)}$:
\begin{equation}\label{for7}
    \alpha_{\rm CO(1-0)} = 4.35 \times (Z/Z_{\odot})^{-1.6} (\rm M_{\odot}pc^{-2}(K\ km\ s^{-1})^{-1})
\end{equation}
where $Z$ is the gas-phase abundance ($12 + \rm log(\rm O/H)$) and $Z_{\odot}$ is the solar value \citep[$\sim$ 8.66;][]{2001ApJ...556L..63A, 2004A&A...417..751A, 2004MNRAS.348L..59P}. \cite{2020ApJ...892..148S} adopted a metallicity-dependent CO-to-H$_2$ conversion factor, $\alpha_{\rm CO}$ (see their formula 4), similar to the metallicity-dependent component of the prescription of \cite{2017MNRAS.470.4750A}, where $Z^{'}$ is the local gas phase metallicity normalized to the solar value appropriate for the \cite{2004MNRAS.348L..59P} metallicity calibration. To apply this prescription uniformly across the PHANGS sample, they estimated the metallicity at one effective radius from the global stellar mass of each galaxy using the mass-metallicity relation of \cite{2019MNRAS.484.3042S}. Then, they extended this prediction to cover the entire galaxy, under the assumption that a universal radial metallicity gradient of $-$0.1 dex/$R_{\rm e}$ \citep[][]{2014A&A...563A..49S}. These estimated radial metallicity profiles then yielded spatially varying values of $\alpha_{\rm CO}$. Subsequently, \cite{2020ApJ...901L...8S} used the same method to investigate the molecular gas properties on $\sim$ 100 pc scales in 70 nearby galaxies. Given that our sample has already measured the global stellar mass \citep[][]{2016ApJS..227....2S, 2018ApJ...859...11S}, we use the same method as theirs to calculate Z. For detailed information, please refer to \cite{2020ApJ...892..148S, 2020ApJ...901L...8S}.

\begin{figure*}[!ht]
 \centering
 \subfigure[]
 {
  \begin{minipage}{0.45\textwidth}
   \centering
   \includegraphics[width=\linewidth]{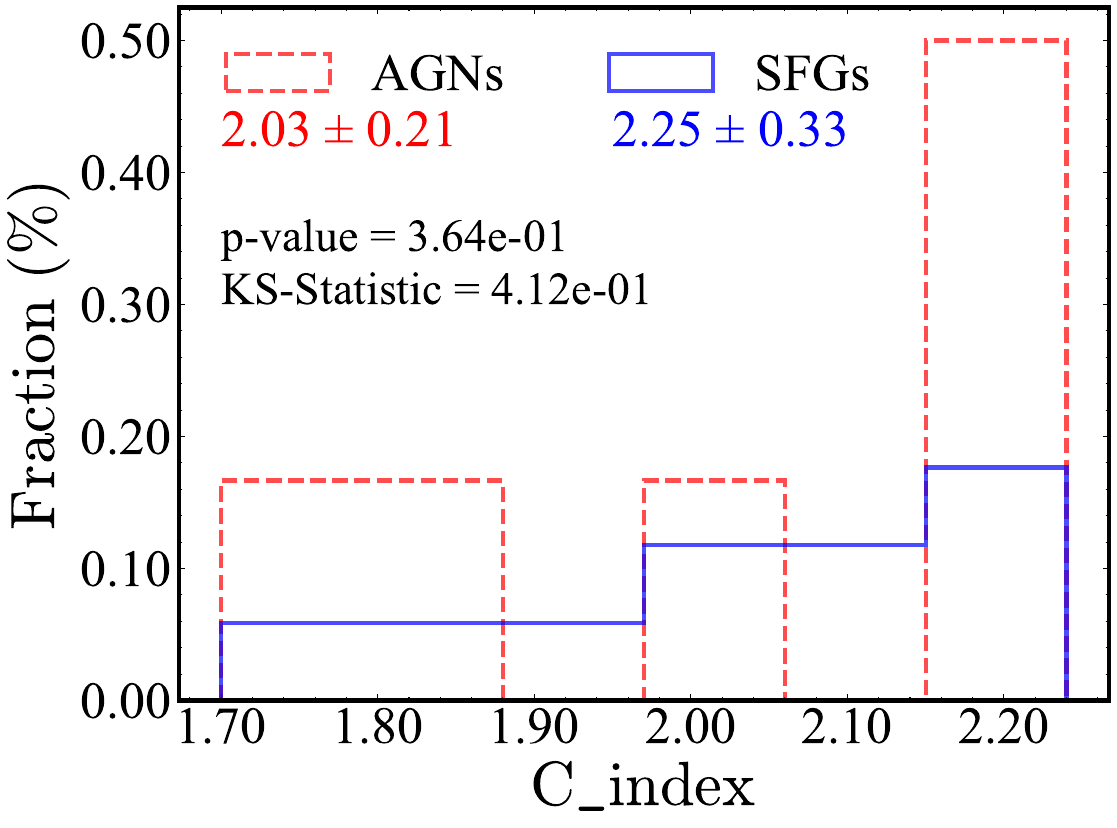}
   \label{C_index}
  \end{minipage}
 }
    \subfigure[]
    {
     \begin{minipage}{0.45\textwidth}
      \centering
      \includegraphics[width=\linewidth]{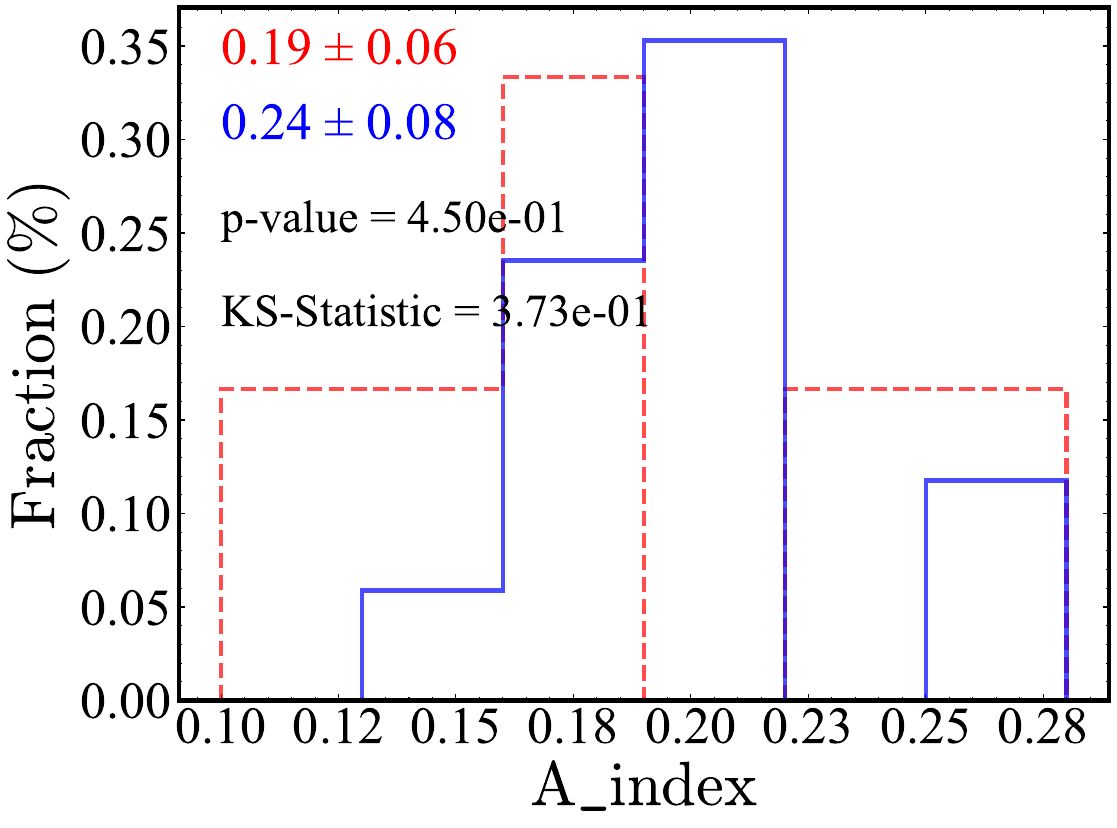}
      \label{A_index}
     \end{minipage}
    } 
    \\
    \subfigure[]
    {
     \begin{minipage}{0.45\textwidth}
      \centering
      \includegraphics[width=\linewidth]{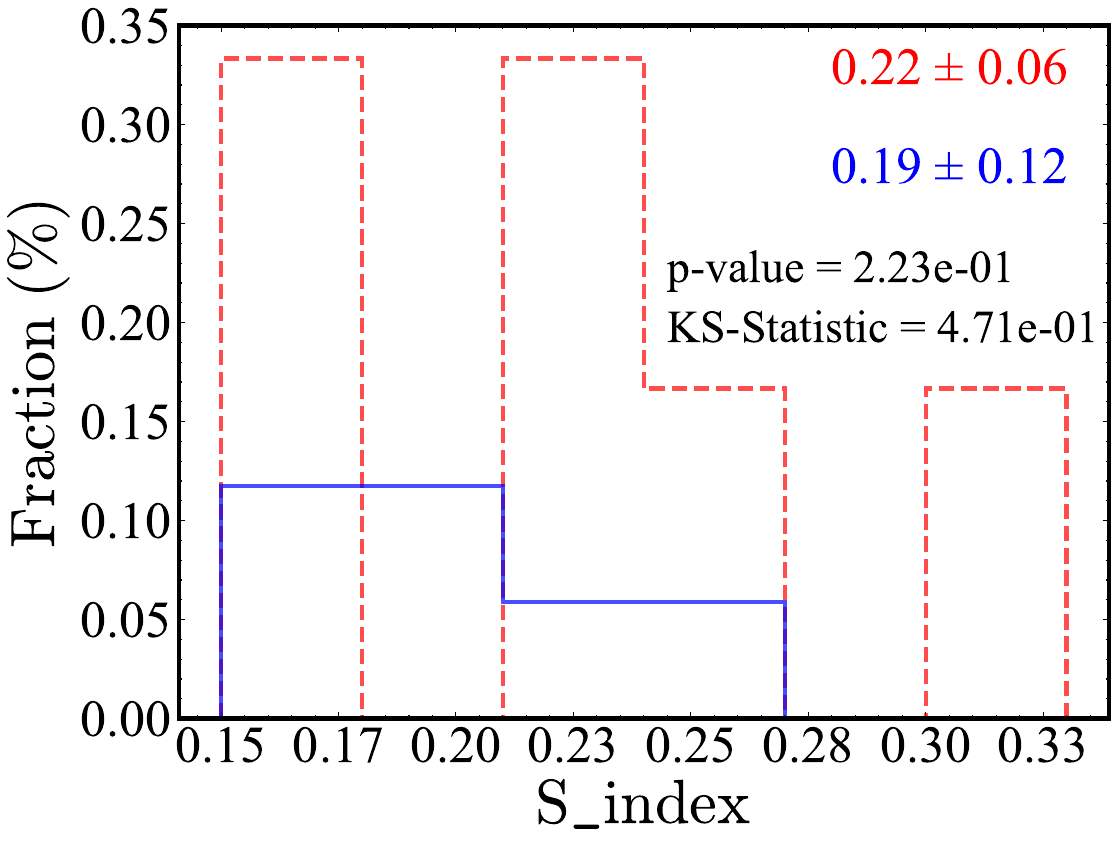}
      \label{S_index}
     \end{minipage}
    }
\caption{CAS parameters of CO(1-0) moment0 map for AGN hosts and SFGs. a),b),c) The histograms of parameters. The black numbers represent the results of the KS test for the parameters of the two samples, while the colored numbers indicate the average values ($\pm$ standard deviation) of the corresponding sample.}
\label{AGN_SFG_CAS}
\end{figure*}

\subsection{Morphological Parameters of CO(1-0) Moment0 Map} \label{sec3.3}

Morphology is a key parameter for studying the properties of AGN hosts, their relationship with AGN, and the formation and evolution of galaxies \citep[e.g.,][]{2003MNRAS.346.1055K, 2012A&A...541A.118P, 2022MNRAS.514..607G}. \citet{2022MNRAS.514..607G} provided the most detailed analysis of how different contributions of AGN to total luminosity may affect six commonly used morphological parameters (e.g., Gini) and the final classification of AGN hosts. Many methods can be used for galaxy classification, including visual classification \citep[e.g.,][]{1926ApJ....64..321H}, parametric methods based on physical and mathematical parameters \citep[e.g.,][]{2010AJ....139.2097P}, and non-parametric methods that measure the different properties of galaxies (e.g., color) related to their morphological types \citep[e.g.,][]{2000ApJ...529..886C, 2019MNRAS.485..452M}.

The definitions of the non-parametric morphological parameters (concentration, asymmetry, smoothness) used in our work are as follows: 

\begin{itemize}
    \item The concentration index (C) is defined as the logarithm of the ratio of circular radii containing 80$\%$ and 20$\%$ of the total flux \citep[][]{2000AJ....119.2645B}:
    \begin{equation}\label{C_index_f}
    C = 5 \times {\rm log}(r_{80}/r_{20}) 
    \end{equation}
    where lower C values correspond to lower fractions of light in the center of the galaxy.

    \item The asymmetry index (A) is calculated by comparing the flux in every pixel using an original map and the rotated one by 180 degrees, and taking into account the background \citep[][]{1994ApJ...432...75A}:
    \begin{equation}\label{A_index_f}
    A = \frac{(\Sigma |I(i,j) - I_{180}(i,j)|) - (\Sigma |B(i,j) - B_{180}(i,j)|)}{2I(i,j)}
    \end{equation}
    where I(i, j) and I$_{180}$(i, j) are the flux in the (i, j) pixel on the original and rotated maps, while B(i, j) and B$_{180}$(i, j) are the background of the original and rotated maps. Highly symmetrical sources, such as undisturbed elliptical/lenticular galaxies, have a smaller A, while galaxies with significant residuals, like spiral galaxies, have a higher A.

    \item The smoothness index (S) is used to measure the small-scale structures (e.g., compact star clusters, star-forming knots) within galaxies and is defined as \citep[][]{2004AJ....128..163L}:
    \begin{equation}\label{S_index_f}
    S = \frac{\Sigma |I(i,j) - I_{\rm S}(i,j)|}{\Sigma |I(i,j)|} - B_{\rm S}
    \end{equation}
    where I$_{\rm S}$ is the original map smoothed by a boxcar of width 0.25r$_{\rm p}$ \citep[r$_{\rm p}$: Petrosian radius;][]{1976ApJ...209L...1P}, and B$_{\rm S}$ is the average smoothness of the background. Higher S indicates a larger number of small-scale structures in a galaxy.
\end{itemize}

In this work, we used the above formula (formulas \ref{C_index_f}, \ref{A_index_f}, \ref{S_index_f}) to calculate the CAS parameters of the CO(1-0) moment0 map.

\subsection{Eddington Ratio}\label{3.4}

As mentioned in Sect. \ref{intro}, AGNs have variable $L_{\rm bol}/L_{\rm Edd}$, which corresponds to different feedback modes. The impact of these modes on cold gas may vary; some cause heating, while others result in ejection \citep[e.g.,][]{2012ARA&A..50..455F, 2022ApJ...938..127X, 2023NatAs...7.1376Z}. Therefore, we need to calculate the $L_{\rm bol}/L_{\rm Edd}$ of our AGNs. Based on the empirical relationship among $M_{\ast}$, $M_{\rm BH}$, and $L_{\rm Edd}$, we can calculate the $M_{\rm BH}$ of the galaxy using its stellar mass \citep[see][]{2023NatAs...7.1376Z}:

\begin{equation}\label{for12}
    {\rm log}\ M_{\rm BH} = (1.81 \pm 0.06) \times {\rm log}\ M_{\ast} + (7.61 \pm 0.50)
\end{equation}
Then, we use the transformed $M_{\rm BH}$ to calculate $L_{\rm Edd}$:
\begin{equation}\label{for13}
    {\rm log}\ L_{\rm Edd} = {\rm log}\ M_{\rm BH} + {\rm log}\ (3.3 \times 10^{4})
\end{equation}

Finally, we obtain $L_{\rm bol}$ of AGN using the empirical relationship \citep[][]{2004ApJ...613..109H}:
\begin{equation}\label{for14}
    L_{\rm bol} = 3500 \times L_{\rm [OIII5007]}
\end{equation}
In Table \ref{tab1}, we present the $L_{\rm bol}/L_{\rm Edd}$ of AGNs, along with the corresponding errors (calculated using formulas \ref{for12}, \ref{for13} and \ref{for14}).

\section{Results} \label{result}
\subsection{Radial Profiles}\label{4.1}

We provide the radial profiles of $D_{\rm n}$4000 along the major axis of galaxies. As shown in Fig. \ref{total_dn4000}, the red squares and blue stars represent the AGN hosts and SFGs, respectively. The colored error bars are the standard deviation of the median, while the gray error bars indicate the 1$\sigma$ error of $D_{\rm n}4000$. The value of each galaxy in each radial bin is determined as the median value of the pixels within that bin. Obtain the median radial profiles by calculating the median of all galaxies of each type \citep[e.g.,][]{2025ApJ...978...59C}.

We find that the radial profile of SFGs exhibits a gradually decreasing trend from the inside out \citep[e.g.,][]{2019ApJ...883L..36H, 2025ApJ...978...59C}, while AGN hosts have different variations. At the center of the galaxy, the AGN hosts and SFGs have similar ages, while the SFGs are significantly younger in the outer regions 1.0 - 1.5$R_{\rm e}$ (lower $D_{\rm n}$4000). As the radial distance increases, the $D_{\rm n}$4000 of the AGN hosts shows a significant decrease at 0.4 - 0.6$R_{\rm e}$ (physical scale: 1 - 2 kpc), followed by a slow increase. The emergence of a young SP in the inner regions of AGN hosts is consistent with previous studies \citep[e.g.,][]{2014A&A...567A.125G, 2019MNRAS.485.3264P, 2021ApJ...913..139G, 2023ApJ...944...30M}. In Sect. \ref{5.1}, we will further discuss the impact of measurement errors.

\subsection{Morphological Parameters}\label{4.2}

In Fig. \ref{AGN_SFG_CAS}, we present a comparison of the CAS parameters of our  AGN sample with the SFGs sample, with the color codes being the same as those in Fig. \ref{total_dn4000}. The black numbers represent the results of the KS test, while the colored numbers indicate the average values ($\pm$ standard deviations) of the corresponding samples. We find that the two samples have similar distributions of CAS parameters, and the KS test also indicates that there is no statistical difference in the distribution of the parameters. This might suggest that the existence of AGN nay not alter the global distribution of molecular gas within the galaxy \citep[e.g.,][]{2020ApJ...899..112S, 2024A&A...690A.140K}.

Observations and simulations have shown that the internal structures of galaxies can lead to changes in the CAS parameters. For instance, the galaxy bar can induce asymmetries (A) and variations in smoothness (S) \citep[e.g.,][]{2015ApJ...802..137C, 2022A&A...661A.105S, 2022MNRAS.514..607G, 2025Galax..13...84G}. Here, we utilize the morphological classification provided by the MaNGA Value-added Catalog - MaNGA Visual Morphologies from SDSS and DESI images\footnote{\url{https://www.sdss4.org/dr17/data_access/value-added-catalogs/?vac_id=manga-visual-morphologies-from-sdss-and-desi-images}} - to determine whether the sample galaxies have bar structures. We find that only one galaxy in the AGN host was identified as a barred galaxy, while 8 galaxies in the SFGs ($\sim$ 47 $\%$) were identified as having a bar structure. However, we find that the CAS of barred AGN (2.03; 0.24; 0.21) is indistinguishable from that of 8 barred SFGs and no-barred AGNs.

Previous studies have demonstrated that AGN activities can have an impact on the measurement of CAS parameters. For instance, \cite{2008ApJ...680..897D} suggest that at higher radio and X-ray luminosities, the hot gas in the host galaxies will be subjected to greater disturbances. Furthermore, the morphological measurements based on optical photometry also revealed possible influences \citep[e.g.,][]{2022MNRAS.514..607G, 2023MNRAS.521.5272T, 2025Galax..13...84G}. Of course, the impact of AGN on CAS might also be rather weak. These results \citep[e.g.,][]{2008ApJ...680..897D, 2022MNRAS.514..607G, 2023MNRAS.521.5272T, 2025Galax..13...84G} are not directly comparable because optical photometry traces stellar light, whereas X-ray and CO observations probe hot and molecular gas, respectively. Their different responses may reflect differences in how AGN energy couples to these components. AGN-driven flows can propagate through low-density channels and disturb diffuse gas, while interacting with only part of the clumpy molecular medium \citep[e.g.,][]{2012ApJ...757..136W}. Such interactions need not produce a detectable change in the overall CO morphologies, even if individual molecular clouds are affected. Optical morphological differences, meanwhile, may involve nuclear emission, stellar population variations, or dust. Our results therefore suggest that the global molecular gas morphology may be comparatively less affected at the scales probed, rather than that molecular gas is immune to feedback. Given our sample of only 6 AGN hosts, this interpretation remains tentative. Furthermore, we are unable to clearly distinguish the effects brought about by the galaxy bar and AGN. Therefore, the absence of differences in the CAS found here might merely be the result of the interplay of various factors, or it could be simply due to the small size of the sample. We will have a discussion on this topic in Sect. \ref{5.2}.

\subsection{Scaling Relations}\label{4.3}
\subsubsection{Spatially rMGMS}

Previous studies have shown a positive correlation between the molecular gas surface density ($\Sigma_{\rm H_{2}}$) and stellar mass surface density \citep[$\Sigma_{\ast}$; e.g.,][]{2013ApJ...777L...4W, 2019ApJ...884L..33L, 2022MNRAS.510.3622B}. In general, stellar mass is believed to reflect the gravitational potential of a galaxy, where molecular gas is formed or retained due to pressure \citep[][]{2019ApJ...884L..33L}. This is called the molecular gas main sequence (MGMS). \citet{2019ApJ...884L..33L} studied the spatially resolved MGMS (rMGMS) of galaxies in the ALMaQUEST survey and found that the local gas mass traces the gravitational potential set by the local stellar mass, or both quantities follow the total mass distributions.

\begin{figure*}
\centering
\includegraphics[width=\linewidth]{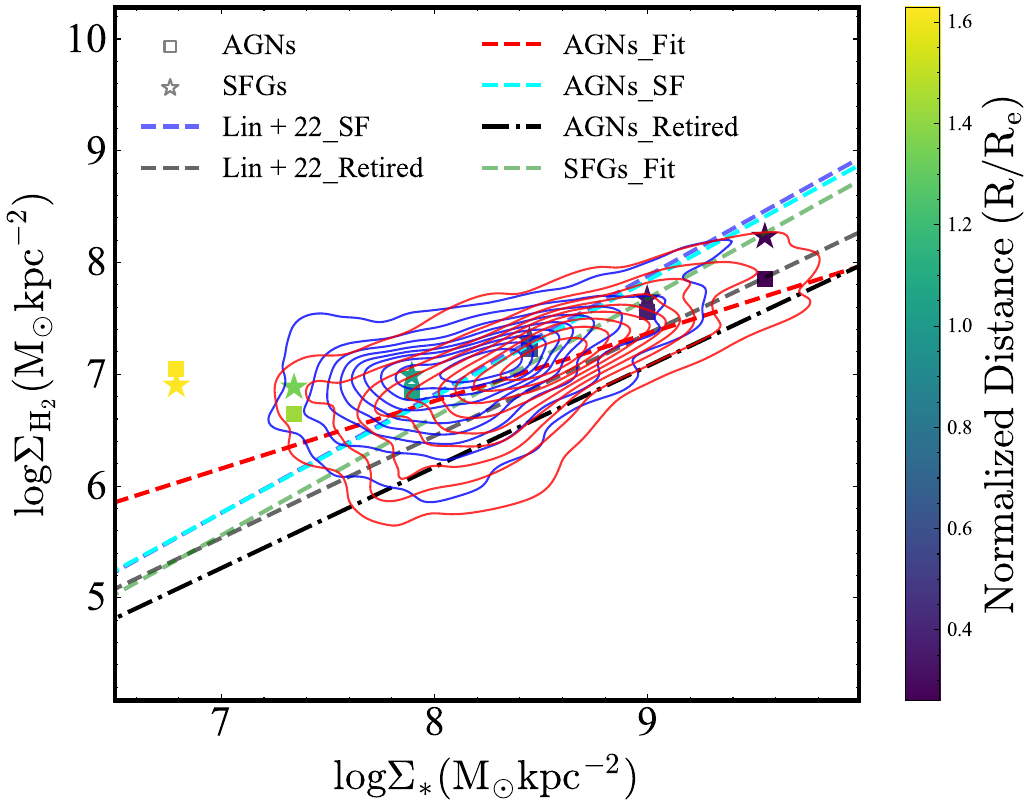}
\caption{The rMGMS. In this figure, the red and green dashed lines represent the best-fitting results for AGNs and SFGs, respectively, while the blue dashed line (black dashed line) come from \citet{2022ApJ...926..175L}, which is the best-fitting result for star-forming (retired) pixels within MS galaxies. The cyan dashed line represents the fitting of the star-forming pixels in our AGNs, while the black dot-dashed line represents the fitting of its retired pixels. The squares (AGN hosts) and stars (SFGs) represent the median in the corresponding $\Sigma_{*}$ bin, and their colors indicate the distance to the center of the galaxy (normalized to $R_{\rm e}$). In addition, the blue and red contours represent the number density distribution of all pixels in SFGs and AGN hosts, respectively.}
\label{spatially rMGMS}
\end{figure*}

Our rMGMS is shown in Fig. \ref{spatially rMGMS}. In this figure, the red (slope: 0.98) and green (slope: 0.82) dashed lines represent the best-fitting result\footnote{The least squares fitting is used to trace the relationship between the molecular gas surface density and stellar mass surface density of galaxies, known as MGMS, and the measurement error of variables is passed as the weight in the fitting process, similar to previous studies \citep[e.g.,][]{2019ApJ...884L..33L, 2024MNRAS.52710201E}.\label{footnote7}} for AGN hosts and SFGs, while the blue dashed line (gray dashed line) come from \citet{2022ApJ...926..175L}, which is the best-fitting result for star-forming (retired) pixels within 28 main-sequence (MS) galaxies in ALMaQUEST (see their Table 1). The squares (AGN hosts) and stars (SFGs) represent the median in the corresponding $\Sigma_{*}$ bin, and their colors indicate the distance to the center of the galaxy (normalized to $R_{\rm e}$). In addition, the blue and red contours represent the number density distributions of all pixels in SFGs and AGNs, respectively. We find that the pixels of AGN hosts and SFGs occupy similar spaces on this plane, and $\Sigma_{\rm H_{2}}$ gradually decreases with increasing distance from the galaxy's center. The best-fitting results between them show differences (red/blue dashed lines in Fig. \ref{spatially rMGMS}), with AGNs having a smaller slope. 

The main difference between the red dashed line and the blue dashed line \citep[][]{2022ApJ...926..175L} in Fig. \ref{spatially rMGMS} is that we do not distinguish the types of pixels (star-forming or retired) in SFGs. Similar to \cite{2022ApJ...926..175L}, we quantify the star-forming pixels and retired pixels in our two samples.Specifically, using the spatially-resolved BPT diagnosis (Fig. \ref{spatially BPT}), we require that star-forming pixels are those in the star-forming regions of the BPT and which have EW$\_$H$\alpha$ > 6\AA. To ensure the number of retired pixels, we only require that their EW$\_$H$\alpha$ < 3\AA, without imposing any restrictions on their positions in the BPT \citep[][]{2022ApJ...926..175L}. By using this method, we identified 3356 star-forming pixels and 135 retired pixels in AGNs, while for SFGs we identified 6088 star-forming pixels and 116 retired pixels respectively. Since the fitting results after excluding the retired pixels in SFGs (slope: 1.04 $\pm$ 0.06; intercept: 6.79 $\pm$ 0.02) are almost identical to the blue dashed line (slope: 1.06 $\pm$ 0.04; intercept: 6.82 $\pm$ 0.02) in Fig. \ref{spatially rMGMS}, we do not provide the fitting results of the star-forming pixels in SFGs here. Finally, we find that the ratio of the two types of pixels (retired/star-forming) in our AGNs obtained through this quantification (135/3356; $\sim$ 4$\%$) is very close to the pixel ratio (108/6122; $\sim$ 2$\%$) given by \cite{2022ApJ...926..175L}. Meanwhile, the fitting of the retired pixels in our AGNs (black dot-dashed lies; slope: 0.90 $\pm$ 0.26) is also consistent with the result (gray dashed line; slope: 0.92 $\pm$ 0.23) of \cite{2022ApJ...926..175L}.

\subsubsection{Spatially rKS}

\begin{figure*}[!htbp]
\centering
\includegraphics[width = \linewidth]{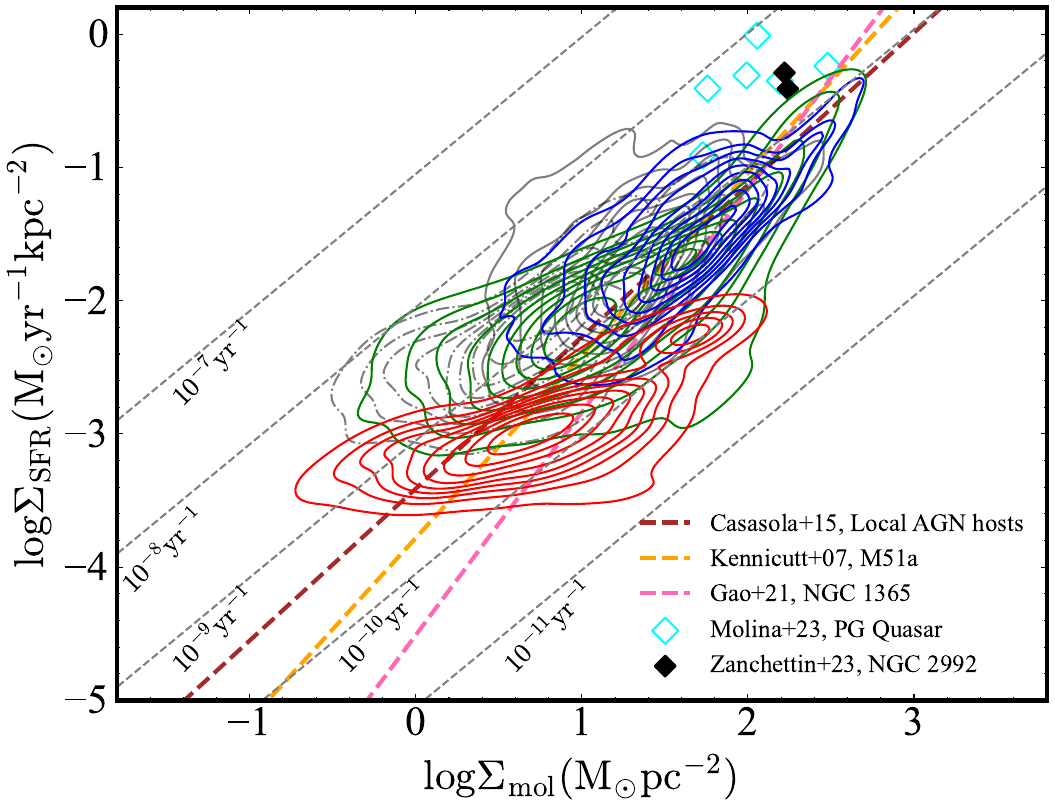}
\caption{The rKS. In this figure, the cyan, and black diamonds represent the PG Quasars from \citet{2023ApJ...944...30M}, NGC 2992 from \citet{2023A&A...679A..88Z}, respectively. The brown, orange, and pink dashed lines are linear fits for the local AGN hosts \citep[][]{2015A&A...577A.135C}, M51a \citep[][]{2007ApJ...671..333K}, and NGC 1365 \citep[$\Sigma_{\rm mol}$ $>$ 1 $M_{\odot}$pc$^{-2}$;][]{2021ApJ...913..139G}, respectively. Furthermore, the gray dashed lines represent different SFEs. The green/blue contours, respectively, represent the distribution of composite/star-forming pixels within AGN hosts, while the red contour represents their AGN-regions pixels. In addition, we use gray contours to represent the star-forming (solid lines) and composite (dashed lines) pixels in the SFGs.}
\label{spatially rKS}
\end{figure*}

The KS law ($\Sigma_{\rm SFR}$ $\varpropto$ $\Sigma_{\rm mol}^{\rm \alpha}$) indicates that molecular gas is the raw material of SF, both for the integrated quantities and also on spatially resolved scales. At spatially resolved scales, rKS quantifies how an increase in resolved $M_{\rm mol}$ results in an increase in SFR \citep[e.g.,][]{2022MNRAS.510.3622B}. In this work, we explore the rKS, as shown in Fig. \ref{spatially rKS}. In this figure, the cyan, and black diamonds represent the PG Quasars from \citet{2023ApJ...944...30M}, two sides of the disk for NGC 2992 from \citet{2023A&A...679A..88Z}, respectively. NGC 2992 is a nearly edge-on spiral galaxy that hosts an AGN known for its extreme variability \citep[e.g.,][]{2023A&A...679A..88Z, 2024ApJ...971..165X}.
The brown, orange, and pink dashed lines are linear fits for the local AGN hosts \citep[][]{2015A&A...577A.135C}, M51a \citep[][]{2007ApJ...671..333K}, and NGC 1365 \citep[$\Sigma_{\rm mol}$ $>$ 1 $M_{\odot}$pc$^{-2}$;][]{2021ApJ...913..139G}, respectively. M51a has a dense molecular disk with a high SFR per unit area, classified as Seyfert 2.0 galaxy \citep[e.g.,][]{2006A&A...455..773V, 2007ApJ...671..333K}, while NGC 1365 is nearby archetype barred spiral Seyfert 1.8 galaxy, which harbors a LLAGN with $L_{\rm bol}$ $\sim$ $2 \times 10^{43}$ ergs$^{-1}$ \citep[e.g.,][]{2021ApJ...913..139G}.
Furthermore, the gray dashed lines represent different constant SFEs. The green/blue contours, respectively, represent the distribution of composite/star-forming pixels within AGN hosts, while the red contour represents their AGN-regions pixels. In addition, we use gray contours to represent the star-forming (solid lines) and composite (dashed lines) pixels in the SFGs.

In general, the distribution of our AGN hosts in the rKS plane is consistent with previous studies \citep[][]{2007ApJ...671..333K, 2015A&A...577A.135C, 2021ApJ...913..139G, 2025PASA...42..140S}. We found that the fit of \citet{2021ApJ...913..139G} is steeper (pink dashed line). This is because their fit only includes pixels with $\Sigma_{\rm mol}$ $>$ 1 $M_{\odot}$pc$^{-2}$ in the galaxy. The study of PG Quasar \citep[][]{2023ApJ...944...30M} revealed the possible existence of AGN positive feedback (cyan diamonds), and their SFRs are determined using rest-frame 230 GHz continuum emission, detected in the central $\leq$ 2 kpc regions of the host galaxies. \mbox{\citet{2023ApJ...944...30M}} pointed out that this emission mainly corresponds spatially to the dust emission produced by the obscured SF activities. Similarly, the fitting of the two sides for NGC 2992 disk (SFR traced by the 6 cm radio continuum) also revealed positive feedback \citep[black diamonds;][]{2023A&A...679A..88Z}. \citet{2024ApJ...971..165X}, however, suggested that there may be two effects of AGN feedback simultaneously in the circumnuclear region for NGC 2992 (SPs fitting). We found that the distribution of Seyfert pixels in AGNs was significantly lower (red contours). Whether compared with previous studies \citep[][]{2007ApJ...671..333K, 2015A&A...577A.135C, 2021ApJ...913..139G, 2023ApJ...944...30M, 2023A&A...679A..88Z}, or with our SFGs and the star-forming/composite pixels of AGN hosts, it showed a lower SFE. In previous studies, in addition to using the continuum to trace SFR, other SFR calibration methods were also employed: H$\alpha$ after dust extinction correction was combined with 24 $\mu$m emission for calibration in \citet{2007ApJ...671..333K}, while in \citet{2015A&A...577A.135C} and \citet{2021ApJ...913..139G}, H$\alpha$ after dust extinction correction was used as a tracer. According to our definition of SFR (Sect. \ref{extinction}), the SFR of AGNs will always be lower than the values given in the literature. Our approach relies on the following assumption: the AGN fraction precisely corrected its contribution to the H$\alpha$ emissions (formula \ref{for4}). Therefore, all the differences between our work and that of previous studies stem from the calibration method for SFR.

\begin{figure*}[!htbp]
 \centering
  \subfigure[]
    {
     \begin{minipage}{0.48\textwidth}
      \centering
      \includegraphics[width=\linewidth]{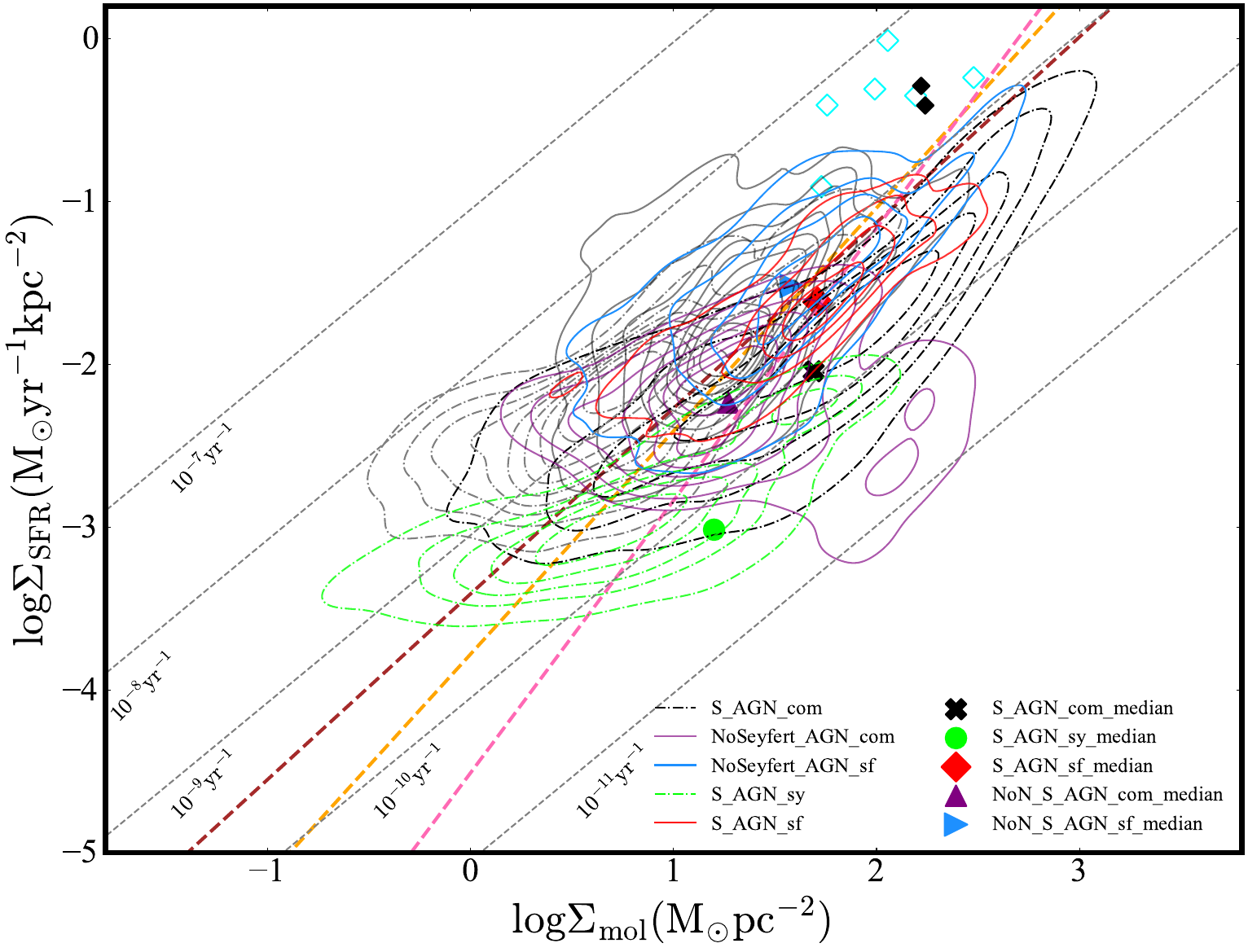}
      \label{spatially_resolved_KS_two_subsamples}
     \end{minipage}
    } 
 \subfigure[]
 {
  \begin{minipage}{0.48\textwidth}
   \centering
   \includegraphics[width=\linewidth]{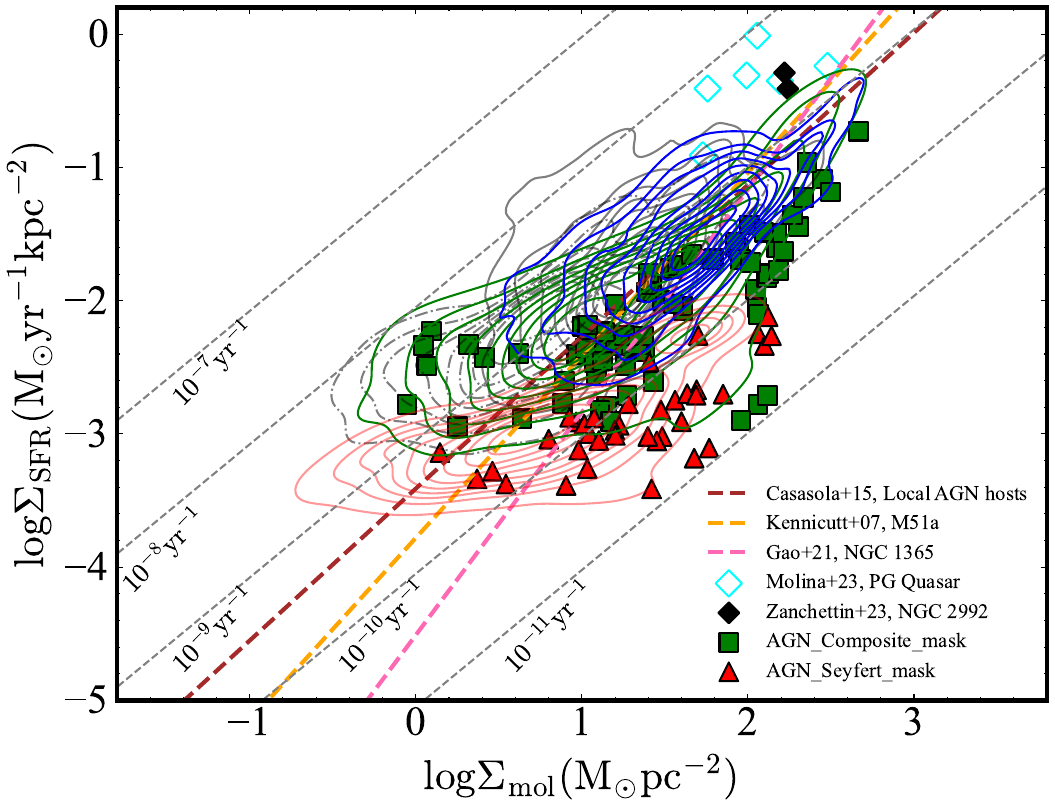}
   \label{spatially_resolved_KS_mask}
  \end{minipage}
 }
    \subfigure[]
    {
     \begin{minipage}{0.48\textwidth}
      \centering
      \includegraphics[width=\linewidth]{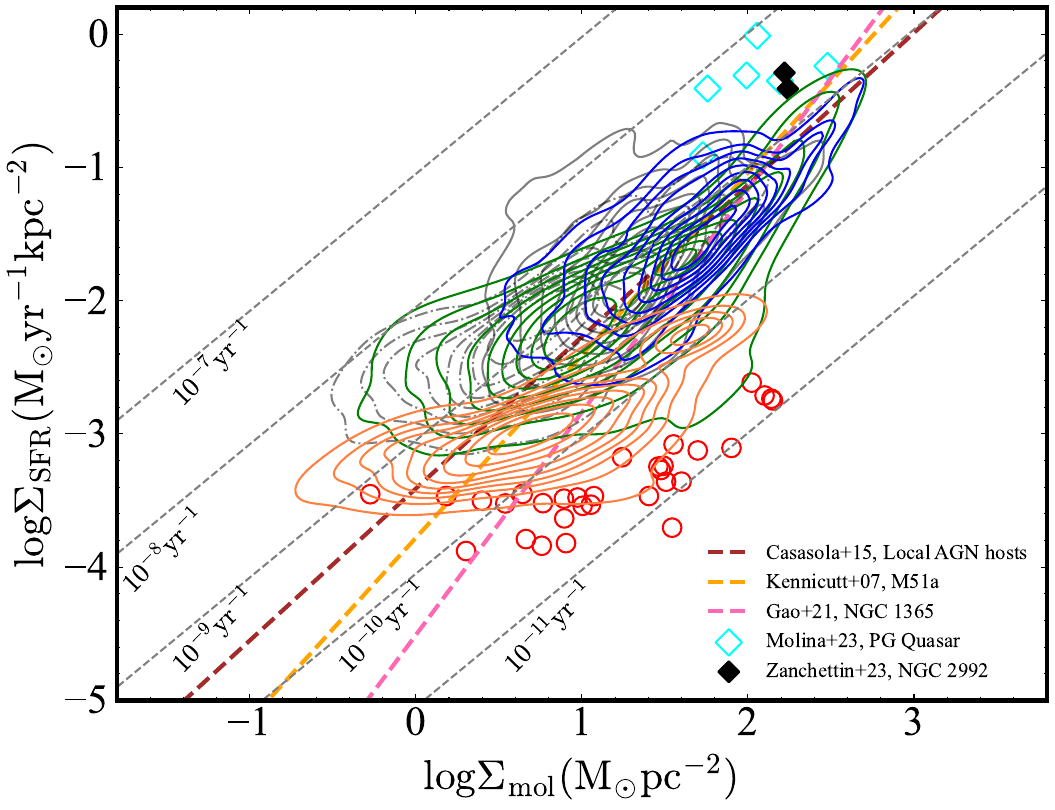}
      \label{spatially_resolved_KS_mask_true}
     \end{minipage}
    } 
   \caption{a) The distribution of the two sub-samples on the rKS. The dodgerblue contours represent the star-forming pixels of NoSeyfert$\_$AGN, and the purple contours represent its Composite pixels. However, the red contours represent the star-forming pixels of Seyfert$\_$AGN, the black dotted contours represent its Composite pixels, and the light-green dotted contours represent its AGN-regions pixels. Furthermore, we have provided the median positions of the data for each contour. b) The rKS after applying the mask. The green solid squares and red solid triangles respectively represent the Composite and AGN-regions pixels within the AGN hosts after masking. c) The true Seyfert pixels are shown as red open circles, while the light-yellow contours represent the number density distributions of pixels in LINER regions. The remaining marks in this figure are exactly the same as those in Fig. \ref{spatially rKS}.}
\label{spatially_resolved_KS_mask_modified}
\end{figure*}

\section{Discussion}\label{discussion}

We compared 6 radio-selected AGN hosts with 17 optically selected SFGs from ALMaQUEST in terms of SPs, non-parametric morphological parameters (CAS) of the CO(1-0) moment0 map, and spatially resolved scaling relations (rMGMS and rKS). 

\subsection{$D_{\rm n}$4000 and Galaxy Internal Structures}\label{5.1}

We found that with increasing radial distance, the $D_{\rm n}4000$ of AGN hosts and SFGs have an opposite change trend (Fig. \ref{total_dn4000}). The former shows a significant decrease at 0.4 - 0.6$R_{\rm e}$ (physical scale: 1 - 2 kpc) before a slow rise, while the latter follows an inside-out quenching mechanism \citep[e.g.,][]{2007ApJ...658.1006M, 2019ApJ...883L..36H, 2024A&A...691A.107C, 2025ApJ...978...59C}. The emergence of young SPs in the inner regions of AGN hosts is consistent with previous studies \citep[][]{2021ApJ...913..139G, 2023ApJ...944...30M, 2024ApJ...971..165X}. The SPs' ages at the center of the two populations are very similar, and both are dominated by relatively old SPs. \citet{2018MNRAS.478.5491M} studied the first 62 AGN hosts from the SDSS-IV MaNGA survey and found that their internal regions ($\leq$ 0.5$R_{\rm e}$) are also dominated by older SPs. As shown in Fig. \ref{total_dn4000}, we extracted the error on $D_{\rm n}4000$ from Pipe3D datacubes and propagated them into the radial profiles. These propagated uncertainties are now shown as gray error bars in Fig. \ref{total_dn4000}. The AGN hosts profile decreases from approximately 1.32 in the innermost radial bin to approximately 1.26 at 0.4 - 0.6$R_{\rm e}$. This decrease is larger than the associated propagated measurement uncertainties, indicating that it is unlikely to arise solely from the measurement errors. Although some studies have shown that the power-law spectrum of AGN may cause contamination to galaxies, resulting in a smaller measured $D_{\rm n}4000$ value, the typical size of contamination is only a few hundred parsecs \citep[$\sim$ 200 pc;][]{2003MNRAS.346.1055K, 2004MNRAS.355..273C, 2017A&A...604A..99C}, which is significantly smaller than the physical size we have obtained here. 

We calculated the average $D_{\rm n}$4000/SFR for each radial bin and then compared it with the average value at the similar radius in SFGs as well as the average value of these galaxies. In Appendix \ref{Dn4000_table}, we present the detailed calculation results (see Table \ref{tab2}). Although AGN hosts have a higher mean $D_{\rm n}4000$ overall than SFGs, the difference depends on radius. At 0.2 - 0.6$R_{\rm e}$, AGN hosts have lower $D_{\rm n}4000$ values than SFGs, consistent with relatively younger SPs, whereas the opposite trend is found at larger radii. For AGN hosts, the low $D_{\rm n}4000$ regions accompanied by a modest increase in SFR ate 0.4 - 0.6$R_{\rm e}$, followed by a progressive decline in SFR as $D_{\rm n}4000$ increases outward. The correspondence between lower Dn4000 and higher SFR suggests locally enhanced recent star formation in AGN hosts. To investigate its association with internal structures, we conducted the following tests. In Appendix \ref{visual overlap}, we have included images to illustrate the relative positions of the $D_{\rm n}4000$ and H$\alpha$ emissions (Fig. \ref{Arc_Sps}). Furthermore, we have overlaid the H$\alpha$ contours with the DECaLS r-band optical images of the galaxies (Fig. \ref{Arc_check}). We found that the dip of $D_{\rm n}4000$ and the peak of H$\alpha$ emissions are precisely related to the spiral arms of galaxies, especially in cases of 8078-6103, 8083-12702, and 8616-9102. Of course, for galaxy 8655-3701 (Fig. \ref{App_C_f6}), it could also be cause by the bar structure within the galaxy. Such dips may exist in the galactic structure of non-AGN galaxies, but our SFG sample is much larger (more diverse), so on average the spiral structure will not show a peak or drop in the radial profile.

\subsection{Multiple Influencing Factors of CO(1-0) CAS}\label{5.2}

We calculated the non-parametric morphological parameters (CAS) of the CO(1-0) moment0 maps. We find that the two samples have similar CAS parameters of CO(1-0) moment0 map, and the KS test also indicated that there is no statistical difference in the distribution of the parameters. \cite{2022MNRAS.512.1522D} found no significant differences in non-parametric molecular gas morphology measures between PHANGS/WISDOM galaxies with and without AGN. Their broader analysis identified effective stellar mass surface density as the strongest predictor of molecular gas morphology, suggesting an important role for the host stellar potential. They proposed that the smoother, more symmetric molecular gas distributions in galaxies with higher stellar mass surface densities may reflect stronger shear, suppressed the formation of stellar features (e.g., arms and bars), and/or reduced gas inflow. In this context, the absence of the morphological difference associated with AGN presence may indicate that host-galaxy structure has a stronger influence on the overall molecular gas distribution, while star formation efficiency and bars contribute to the scatter. Our results are compatible with this interpretation, although our AGN sample does not allow us to establish the dominant mechanism. Previous studies have indicated that the resolution can have a significant impact on the CAS parameters \citep[e.g.,][]{2016MNRAS.461.1656G, 2021MNRAS.507..886T, 2024MNRAS.528.5558W}. In \cite{2022MNRAS.512.1522D}, they utilized the higher-resolution data, and reported that  the morphological parameters of AGNs are largely indistinguishable from those of normal galaxies. These results suggest that the presence of AGN activity alone does not necessarily produce measurable differences in the global morphological properties traced by CAS, and the similarities between morphological of AGNs and non-AGNs are consistent regardless of resolution.

One possible concern is the influence of bar structures, which have been shown to affect galaxy morphology by redistributing gas and stars and enhancing central mass concentration \citep[e.g.,][]{2025Galax..13...84G, 2025ApJS..279...16G}. In particular, unresolved or poorly resolved bars may introduce biases in CAS measurements, as their contribution to the central light distribution can be smeared by limited spatial resolution \citep[e.g.,][]{2018MNRAS.474.5372E}. However, only one galaxy in our AGNs is classified as barred, indicating that bars are unlikely to be a dominant factor shaping the overall CAS distribution of the AGN population studied here. Therefore, the absence of significant CAS differences between AGNs and SFGs is unlikely to be driven by the presence of resolved bars. Nevertheless, we cannot completely exclude the possibility that small-scale or unresolved bar-like structures remain undetected in the available data and contribute modestly to the measured structural parameters \citep[e.g.,][]{2003ApJS..147....1C, 2007ApJ...659..162T}.

Furthermore, different growth histories can also lead to different CAS. According to \cite{2022MNRAS.516.1462T}, we find that in our AGN sample, two sources (8081-9101, 8615-3703) were classified as mergers or post-merger, while in our SFG sample, three sources (8084-3702, 8156-3701, 8450-6102) were classified in this manner. We find that AGNs exhibiting merger or post-merger signatures generally show systematically higher CAS parameters than non-interacting AGNs. The average CAS values of the former are 2.13 (C), 0.21 (A) and 0.27 (S), while those of non-interacting AGNs are 1.97, 0.185 and 0.19. This trend is consistent with previous observational and theoretical studies demonstrating that galaxy interactions and mergers can significantly reshape galaxy morphology. Gravitational perturbations during mergers induce tidal features, asymmetric stellar distributions, and enhanced star-forming clumps, leading to elevated A and S \citep[e.g.,][]{2000ApJ...529..886C, 2003ApJS..147....1C, 2008ApJ...672..177L, 2010MNRAS.404..590L}. At the same time, merger-driven gas inflows efficiently transport gas toward the central regions, promoting both bulge growth and BH accretion, which can increase the C \citep[e.g.,][]{2008ApJ...672..177L}. Therefore, the systematically higher CAS values observed in our merger and post-merger AGNs likely reflect the combined effects of merger-induced structural disturbances, enhanced central SF, and the buildup of dense central components \citep[][]{2000ApJ...529..886C, 2003ApJS..147....1C, 2008ApJ...672..177L, 2010MNRAS.404..590L}. 

Because the merger history can significantly enhance the asymmetry and clumpiness of the molecular gas distribution, resulting in higher CAS parameters, the different merger ratios between AGN and SFG samples may have a significant impact on the overall CAS comparison of them. We found that the proportion of galaxies with merger history in AGN hosts (2/6, $\sim$ 33 $\%$) is approximately twice that of SFGs (3/17, $\sim$ 18 $\%$). This indicates that the high CAS systems driven by mergers in the AGN sample are more common. The morphological perturbations caused by mergers may enhance or even dominate the CAS differences between AGNs and SFGs. Therefore, when interpreting the molecular gas morphology of AGN hosts, the merger history needs to be considered as an important confounding factor. Here, we did not quantify the impact of mergers on CAS, but instead attempted to understand how mergers might shape the results we presented. Although the AGN sample has a higher merger fraction and the merged hosts have higher A and C values, overall the AGN sample has lower A and C values than the SFG sample (see Fig. \ref{AGN_SFG_CAS}).

\subsection{Representation of Scaling Relations}\label{5.3}

We investigated the spatially resolved scaling relations (rMGMS and rKS) of the two populations. In Fig. \ref{spatially rMGMS}, the best fit of all pixels in the AGN hosts is also close to the result of \citet{2022ApJ...926..175L}, although there is a deviation at the low-mass end. We also found that the pixels of AGN hosts occupy the same space as SFGs in this plane. These results suggest that the gas fraction of each pixel in AGN hosts is comparable to that in SFGs. \citet{2019ApJ...884L..33L} found that pixels with a higher sSFR tend to be located at the upper end of the rMGMS, indicating that SF is increased in regions with enhanced gas fraction. Subsequently, the influence of the gas fraction on the sSFR was further discussed in \citet{2020MNRAS.492.6027E}. In Fig. \ref{spatially rMGMS}, the difference between our best fit of the pixels in SFGs and \citet{2022ApJ...926..175L} is that we do not distinguish the types of pixels (star-forming or retired). In addition, both samples show that the $\Sigma_{\rm H_{2}}$ gradually decreases with increasing distance from the galaxy's center. We used the same method to classify the pixels (star-forming or retired) and obtained results similar to those of \citet{2022ApJ...926..175L}. This indicates that the selected galaxies are more inclined towards star formation, even for those with AGNs. This AGN classification method provides pixels that are consistent with star formation, at least for rMGMS. In particular, the radio-selected AGNs may imply the past AGN activities that have not led to an excess retired pixels.

The most interesting thing is that we found that the AGN-regions pixels within AGN hosts are significantly lower than SFGs and their other two types of pixels in the rKS plane (Fig. \ref{spatially rKS}), namely the lower SFE. The AGN-regions pixels in AGN hosts are located in the outer regions of the galaxy, and the SPs' ages of the outer regions ($D_{\rm n}4000$) within AGN hosts are significantly older compared to SFGs. Furthermore, we calculated the average $D_{\rm n}4000$ for the three regions (SF: 1.25 $\pm$ 0.01; Composite: 1.27 $\pm$ 0.01; AGN: 1.31 $\pm$ 0.02) in AGNs and found that the AGN regions exhibited a significantly larger average (i.e. older SPs). All pixels in AGN hosts occupied positions on the rMGMS that were similar to those of SFGs, meaning there were comparable gas fractions (Fig. \ref{spatially rMGMS}). However, the AGN-regions pixels within AGN hosts exhibit lower SF levels (Fig. \ref{spatially rKS}). We suggest that there is negative AGN feedback, which suppresses SF \citep[e.g.,][]{2006ApJS..163....1H, 2012ARA&A..50..455F, 2019MNRAS.486.2827D, 2023ApJ...944...30M}. The comparable molecular gas content but lower SFE in the Seyfert-classified regions is consistent with a possible form of AGN negative feedback that reduces the efficiency of converting molecular gas into stars without substantially depleting the gas reservoir. For example, AGN-driven turbulence could inhibit the formation of gravitationally bound structures while leaving much of the CO-traced molecular gas in place \citep[3C 326N;][]{2015A&A...574A..32G}. According to the spatially-resolved BPT diagnosis (Fig.\ref{spatially BPT} and Fig. \ref{remainming BPT}), among our six AGNs, only three showed optical feature-based Seyfert emission (hereafter Seyfert$\_$AGN), while the remaining three did not (NoSeyfert$\_$AGN). By comparing, we found that the molecular gas content (or fraction, f$_{\rm H2}$) of Seyfert$\_$AGN is comparable to that of NoSeyfert$\_$AGN, and both contain only one galaxy identified as a merger or post-merger. 

However, by using the morphological fitting parameter S$\acute{\rm e}$rsic index provided by the NSA catalog$^{\ref{NSA}}$, we found that the S$\acute{\rm e}$rsic index of the bulge in Seyfert$\_$AGN (average: 2.45) is higher than that in NoSeyfert$\_$AGN ($\sim$ 2.0). In Fig. \ref{spatially_resolved_KS_two_subsamples}, we have presented the distribution of Seyfert$\_$AGN and NoSeyfert$\_$AGN on rKS. In the figure, the dodgerblue contours represent the star-forming pixels of NoSeyfert$\_$AGN, and the purple contours represent its Composite pixels. However, the red contours represent the star-forming pixels of Seyfert$\_$AGN, the black dotted contours represent its Composite pixels, and the light-green dotted contours represent its AGN-regions pixels. We found that, apart from the AGN-regions pixels (light-green contours in Fig. \ref{spatially_resolved_KS_two_subsamples}), the remaining two pixel types in these two sub-samples had similar distributions within the rKS plane (orange dotted contours - light-yellow contours; black contours - green dotted contours). Considering that Seyfert$\_$AGN has a larger S$\acute{\rm e}$rsic index, we suspect that the AGN negative feedback observed here might be entirely the result of morphological quenching caused by the bulge. Or, the relatively low SFE observed in Seyfert$\_$AGN might be due to the result of the combined influence of morphological quenching and AGN feedback. Furthermore, the comparison between Seyfert$\_$AGN and NoSeyfert$\_$AGN (Fig. \ref{spatially_resolved_KS_two_subsamples}), indicates that the SFR in these systems is not always suppressed (orange dotted contours - light-yellow contours; black contours - green dotted contours). This implies that the kinetic mode does not always have the power and potential to globally suppress SF. The distributions of the two pixel types (orange dotted contours and green dotted contours) in rKS (Fig. \ref{spatially_resolved_KS_two_subsamples}) within the NoSeyfert$\_$AGN indicate that although we observed negative feedback in some galaxies (Seyfert$\_$AGN), such phenomenon does not exist in all of our AGN hosts. It is worth noting that the SFR calculated through our methodology will be lower than that of NoSeyfert$\_$AGN. Therefore, it is necessary to find a suitable alternative tracer for SFR (e.g., SSP derived SFR, UV, or IR) to confirm our results. For example, using the method of stellar population fitting, piXedfit can provide spatially-resolved spectral energy distributions of galaxies \citep[][]{2021ApJS..254...15A}, including the distribution of SFR. Although these methods have their own biases in AGN studies, they can provide external confirmation of these feedback signatures.

As mentioned in Sect. \ref{intro}, AGN feedback has different modes (radiative or kinetic mode), and they have different effects on gas. In table \ref{tab1}, we present the Eddington ratio (L$_{\rm bol}$/L$_{\rm Edd}$) of AGNs. Based on previous studies \citep[e.g.,][]{2014ARA&A..52..589H}, we found that the Eddington ratios of these galaxies are all less than -3 (log$L_{\rm bol}/L_{\rm Edd}$), which indicates that they are LLAGNs. We found that based on their Eddington ratios, these AGNs would be in kinetic mode \citep[e.g.,][]{2012ARA&A..50..455F, 2014ARA&A..52..589H}. Therefore, based solely on the samples we currently use, different feedback modes should not cause any changes in the distribution of galaxies in terms of their scaling relations (rMGMS and rKS).

In Sect. \ref{extinction}, we corrected the AGN contamination for the emission of H$\alpha$ based on \cite{2020MNRAS.499.5749J} and \cite{2021ApJ...923....6J}. However, since the SFR estimates strongly depend on the AGN-removal procedure described above, additional checks are required to assess the robustness of our results. To this end, we filter the pixels based on their $f_{\rm AGN}$. Specifically, pixels in which the AGN contribution exceeds 70$\%$ are excluded from the rKS analysis. The rKS obtained after applying the mask is shown in Fig. \ref{spatially_resolved_KS_mask}. In this figure, the green solid squares and red solid triangles respectively represent the Composite and AGN-regions pixels within AGN hosts after masking, and the remaining masks are exactly the same as those in Fig. \ref{spatially rKS}. We found that the AGN-regions pixels after masking (red solid triangles in Fig. \ref{spatially_resolved_KS_mask}) still exhibited a significantly lower SFE. According to the method described in Sect. \ref{extinction}, the SFR values of these pixels obtained are actually the upper limits. The actual values may be even lower than those shown in Fig. \ref{spatially_resolved_KS_mask}. Therefore, we believe that the impact brought by this correction method can be disregarded.

Furthermore, through the spatially-resolved BPT diagnosis, we found that the majority of the AGN-regions pixels fall within the LINER regions. Therefore, we conducted a simple examination to prove that the ionization from AGN is actually tied to the negative SF feedback. As shown in Fig. \ref{spatially_resolved_KS_mask_true}, we use the orange contours represent the number density distributions of pixels that fall within the LINER regions of the BPT diagram, and the red open circles are the true Seyfert pixels. 
Although Seyfert pixels constitute only a small fraction of the AGN-region pixels, they generally exhibit lower SFE than LINER pixels. This distinction is important because shocks can contribute to LINER-like emission \citep[e.g.,][]{2019MNRAS.487.4153D, 2025ApJ...988..261Z}, so the emission from LINER pixels cannot necessarily be attributed entirely to AGN photoionization \citep[e.g.,][]{2025ApJ...988..261Z}. Consequently, the rKS distribution of the combined AGN-region pixel (red contours in Fig. \ref{spatially rKS}) may mix regions with different excitation mechanisms. By separating the LINER pixels, we find that the lower SFE persists in the true Seyfert subset (red open circles in Fig. \ref{spatially_resolved_KS_mask_true}), where AGN photoionization is more directly indicated and the concern about shocks producing LINER-like emission is less relevant. This result supports a possible connection between AGN activities and suppressed star formation. Distinguishing Seyfert from LINER pixels is therefore essential when interpreting the rKS distribution as a potential signature of AGN feedback.

\section{Conclusions}\label{conclusions}

We selected 6 radio-selected AGN hosts from ALMaQUEST, and compared them with SFGs based on the SPs, non-parametric morphological parameters (CAS) of the CO(1-0) moment0 map, and classic scaling relations (rMGMS and rKS). We attempted to understand the influence of AGN feedback on kpc scales for their hosts. The key outcomes of our investigation are outlined below.

\begin{itemize}
    \item We found that AGN hosts and SFGs have completely different age gradients in SPs ($D_{\rm n}4000$), with the former showing a significant decrease followed by a slow increase as the radial distance increases, while SFGs follow the inside-out quenching mechanism. By comparing the $D_{\rm n}$4000, H$\alpha$ emissions with the optical images of AGN hosts, we suggest that the decrease of $D_{\rm n}$4000 is caused by the internal structures of galaxies -- spiral arms.
    
    \item We found that the two samples have similar CAS parameters, and the KS test also indicated that there was no statistical difference in their distributions. Based on previous studies, the CAS parameters of CO(1-0) may be influenced by a combination of factors such as galaxy bars, AGN, galaxy merger, and the resolution. After comparing with previous studies \citep[][]{2022MNRAS.512.1522D}, we suggest that our results might be the result of the combined effect of unresolved bars in galaxies and the merger of galaxies. However, due to the limited size of our sample, we are unable to provide a definite conclusion at present.

    \item All pixels of AGN hosts and SFGs occupy the same space on the rMGMS, and their best-fitting results show slight differences. This indicates that the gas fractions of pixels in the two populations are comparable. Furthermore, we found that $\Sigma_{\rm H_{2}}$ decreases with increasing radial distance in two samples. After classifying the pixels in the galaxy (star-forming or retired), both the ratios of these two types of pixels, as well as the fitting results (Fig. \ref{spatially rMGMS}), are very similar to  \citet{2022ApJ...926..175L}. The low proportion of retired pixels and their distribution on the rMGMS (traced by the black dotted line in Fig. \ref{spatially rMGMS}) indicate that in our AGN hosts with a fixed stellar mass surface density, the CO detection regions do not show a molecular gas deficit. Given that our AGNs are selected based on their radio observations, this might imply that the past AGN activity did not result in an excess of retired pixels.

    \item For AGN hosts, the Seyfert pixels occupy significantly lower positions in the rKS plane than their other two types of pixels. Their lower SFE may imply negative AGN feedback, which suppresses their SF. The comparable molecular gas content but lower SFE in the Seyfert-classified regions is consistent with a possible form of AGN negative feedback that reduces the efficiency of converting molecular gas into stars without substantially depleting the gas reservoir. For example, AGN-driven turbulence could inhibit the formation of gravitationally bound structures while leaving much of the CO-traced molecular gas in place \citep[][]{2015A&A...574A..32G}. Furthermore, Seyfert$\_$AGNs have a relatively large  S$\acute{\rm e}$rsic index, so we have reason to suspect that the AGN feedback observed here might also be the result of the galaxy morphological quenching.
\end{itemize}

The AGN hosts analyzed in this study are found to be in the kinetic mode, suggesting that our conclusions primarily reflect the influence of kinetic-mode AGNs on the SF activities. Extending this analysis to larger samples that include AGN spanning different accretion and feedback states represents a promising avenue for future work. A systematic comparison between radiative-mode and kinetic-mode AGNs therefore has the potential to uncover the physical mechanisms through which AGN feedback regulates SF and shapes the co-evolution of SMBHs and their host galaxies. All of our results are for samples selected from the ALMaQUEST survey, and it is worth noting that, considering the size of our samples, these results should be used with caution.

\vspace{5mm}

\section*{Acknowledgements}
We are grateful to the anonymous referee for her/his thoughtful review and very constructive suggestions, which greatly improved this paper.
We thank Dr. Qiusheng Gu, Dr. Feng Yuan, Dr. Yinghe Zhao, Dr. Yan-Mei Chen, Dr. Cai-Na Hao, Dr. Min Du and Dr. Zhi Li for helpful discussions. 
We acknowledge the National SKA Program of China, No. 2025SKA0150103, the National Key R$\&$D Program of China (Grant NO.2023YFA1607904), the National Natural Science Foundation of China (NSFC) grant 12221003, and the science research grants from the China Manned Space Project with No. CMS-CSST-2021-A04, CMS-CSST-2025-A07 and No. CMS-CSST-2025-A10.
P.C. acknowledges the Natural Science Foundation of Sichuan Province (No. 2026NSFSC0807).
J.W. acknowledges the NSFC 12333002, 12633004, 12192220.
T.F. is supported by the NSFC under Nos. 11890692, 12133008, 12550002.
X.X. acknowledges the NSFC grant 12403018, the China Postdoctoral Science Foundation (No. 2023M741639), and the Jiangsu Funding Program for Excellent Postdoctoral Talent (No. 2024ZB249). 
X.C. acknowledges support from the China Postdoctoral Science Foundation under Grant Number 2025M773188 and the Postdoctoral Fellowship Program of CPSF under Grant Number GZB20250734. 
Qingshun Hu acknowledges the Natural Science Foundation of Sichuan Province (No. 2025ZNSFSC0879).

This research made use of Photutils, an Astropy package for detection and photometry of astronomical sources \citep[][]{2022zndo...7419741B}. This paper makes use of the following data from ALMA: 
ADS/JAO.ALMA\#2015.1.01225.S,
ADS/JAO.ALMA\#2017.1.01093.S,
ADS/JAO.ALMA\#2018.1.00541.S, ADS/JAO.ALMA\#2018.1.00558.S. 
ALMA is a partnership of ESO (representing its member states), NSF (USA) and NINS (Japan), together with NRC (Canada), MOST and ASIAA (Taiwan), and KASI (Republic of Korea), in cooperation with the Republic of Chile. The Joint ALMA Observatory is operated by ESO, AUI/NRAO, and NAOJ. The National Radio Astronomy Observatory is a facility of the National Science Foundation operated under cooperative agreement by Associated Universities, Inc.

This survey makes use of the MaNGA survey, part of the Sloan Digital Sky Survey (SDSS). Funding for the Sloan Digital Sky Survey IV has been provided by the Alfred P. Sloan Foundation, the U.S. Department of Energy Office of Science, and the Participating Institutions. SDSS-IV acknowledges support and resources from the Center for High-Performance Computing at the University of Utah. The SDSS website is \url{www.sdss.org}. SDSS-IV is managed by the Astrophysical Research Consortium for the Participating Institutions of the SDSS Collaboration including the Brazilian Participation Group, the Carnegie Institution for Science, Carnegie Mellon University, the Chilean Participation Group, the French Participation Group, Harvard-Smithsonian Center for Astrophysics, Instituto de Astrof\'{i}sica de Canarias, The Johns Hopkins University, Kavli Institute for the Physics and Mathematics of the Universe (IPMU) / University of Tokyo, Lawrence Berkeley National Laboratory, Leibniz Institut f\"{u}r Astrophysik Potsdam (AIP), Max-Planck-Institut f\"{u}r Astronomie (MPIA Heidelberg), Max-Planck-Institut f\"{u}r Astrophysik (MPA Garching), Max-Planck-Institut f\"{u}r Extraterrestrische Physik (MPE), National Astronomical Observatory of China, New Mexico State University, New York University, University of Notre Dame, Observat\'{o}rio Nacional / MCTI, The Ohio State University, Pennsylvania State University, Shanghai Astronomical Observatory, United Kingdom Participation Group, Universidad Nacional Aut\'{o}noma de M\'{e}xico, University of Arizona, University of Colorado Boulder, University of Oxford, University of Portsmouth, University of Utah, University of Virginia, University of Washington, University of Wisconsin, Vanderbilt University, and Yale University.
\vspace{5mm}

Facilities: Sloan, ALMA

Software: astropy \citet[][]{2013A&A...558A..33A, 2018AJ....156..123A, 2022ApJ...935..167A}; Tool for OPerations on Catalogues And Tables \citep[TOPCAT;][]{2005ASPC..347...29T} ; DS9 \citep[][]{2003ASPC..295..489J}




\bibliography{main}
\bibliographystyle{aa}

\onecolumn
\begin{appendix}
\section{Spatially Resolved BPT Diagrams for Remaining AGN Hosts}\label{remaining BPT}

\begin{figure*}[!ht]
 \centering
 \subfigure[]
 {
  \begin{minipage}{0.60\textwidth}
   \centering
   \includegraphics[width=\linewidth]{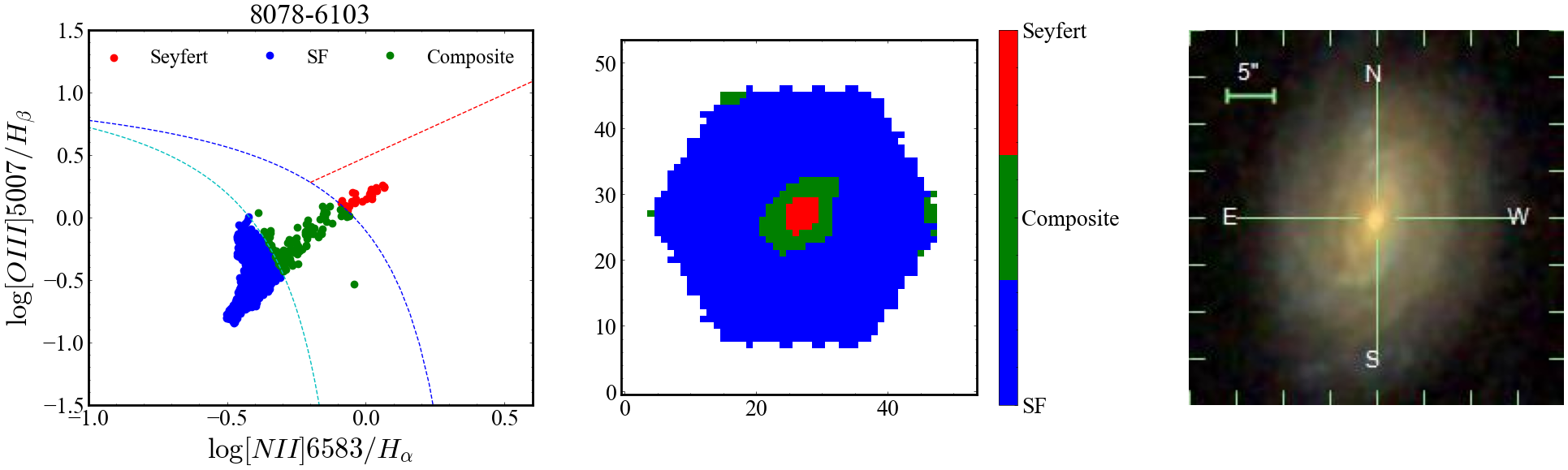}
   \label{App_f1}
  \end{minipage}
 }
    \subfigure[]
    {
     \begin{minipage}{0.60\textwidth}
      \centering
      \includegraphics[width=\linewidth]{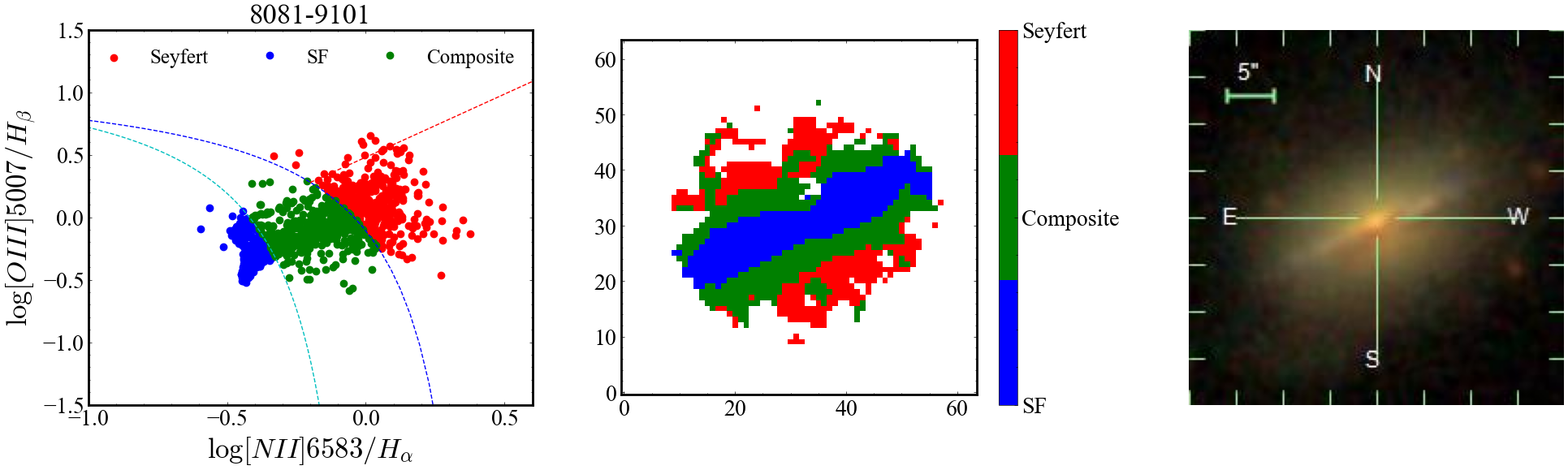}
      \label{App_f2}
     \end{minipage}
    }
    \\
    \subfigure[]
    {
     \begin{minipage}{0.60\textwidth}
      \centering
      \includegraphics[width=\linewidth]{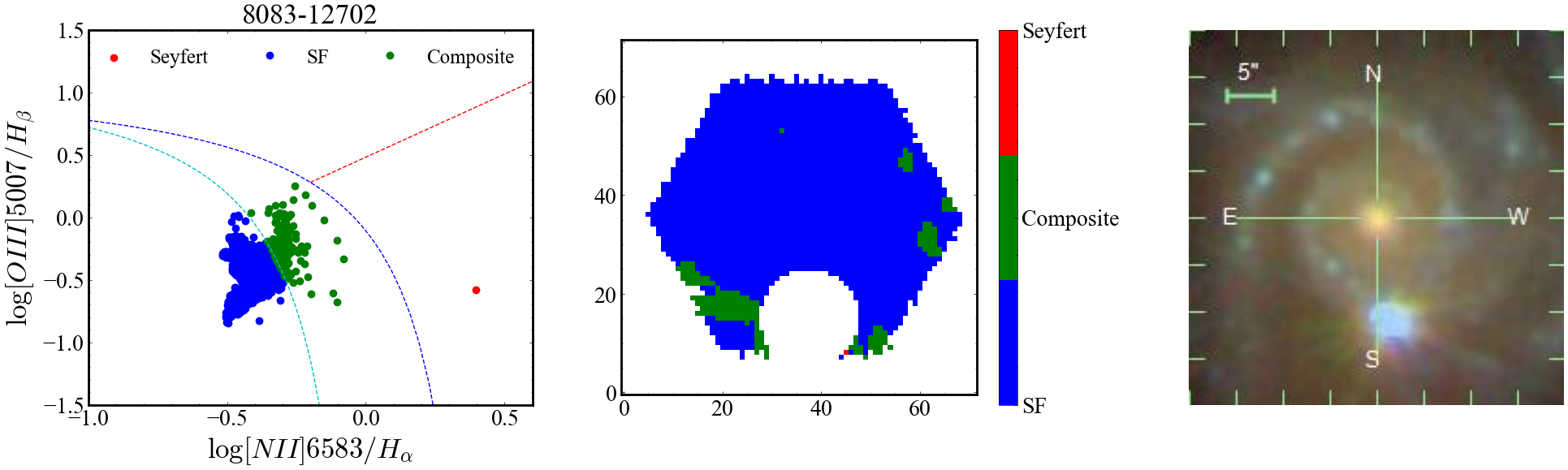}
      \label{App_f3}
     \end{minipage}
    }
    \subfigure[]
    {
     \begin{minipage}{0.60\textwidth}
      \centering
      \includegraphics[width=\linewidth]{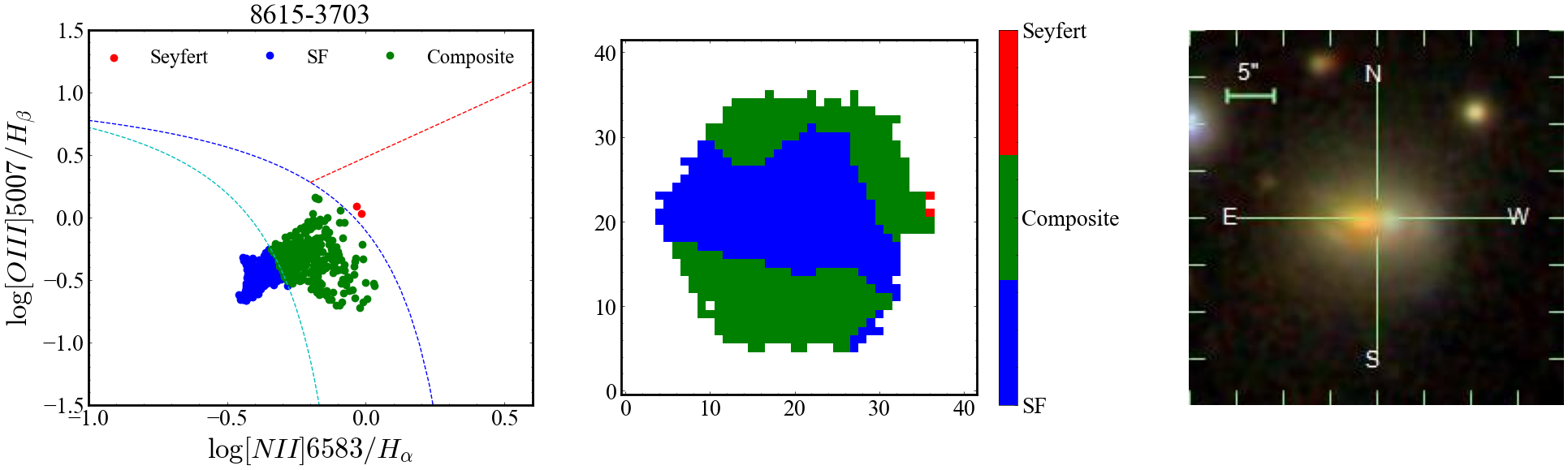}
      \label{App_f4}
     \end{minipage}
    }
    \\
    \subfigure[]
    {
     \begin{minipage}{0.60\textwidth}
      \centering
      \includegraphics[width=\linewidth]{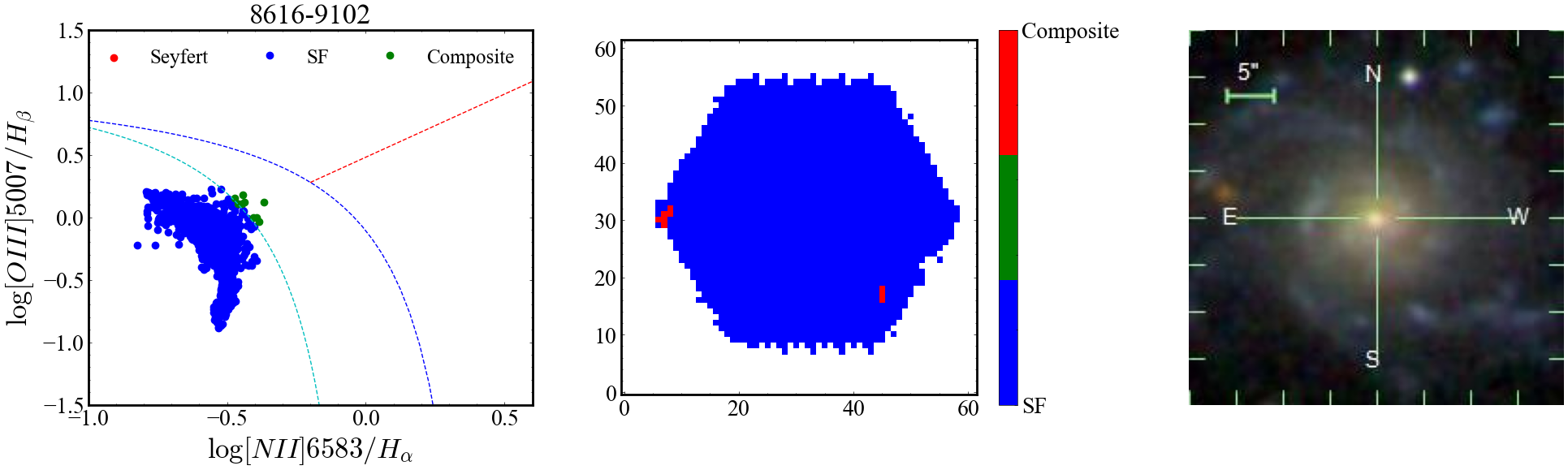}
      \label{App_f5}
     \end{minipage}
    }
\caption{The spatially resolved BPT diagrams for remaining AGN hosts. The markings in this figure are the same as those in Fig. \ref{spatially BPT}.}
\label{remainming BPT}
\end{figure*}

\section{The average $D_{\rm n}$4000/SFR of  each ring, as well as the average values for these galaxy}\label{Dn4000_table}

\begin{table}[h!]
\small
\begin{center}
\caption{Average $D_{\rm n}$4000 and log SFR ($M_{\odot}$yr$^{-1}$)}\label{tab2}
\begin{tabular}{ccccccccc}
\hline\hline
Sample & Average Values & 0.0-0.2$R_{\rm e}$ & 0.2-0.4$R_{\rm e}$ & 0.4-0.6$R_{\rm e}$ & 0.6-0.8$R_{\rm e}$ & 0.8-1.0$R_{\rm e}$ & 1.0-1.2$R_{\rm e}$ & 1.2-1.4$R_{\rm e}$\\
\hline
$D_{\rm n}4000$$_{\rm AGNs}$ & 1.28 $\pm$ 0.02 & 1.37 $\pm$ 0.008 &  1.265 $\pm$ 0.015 & 1.27 $\pm$ 0.022 & 1.31 $\pm$ 0.025 & 1.32 $\pm$ 0.023 & 1.34 $\pm$ 0.022 & 1.36 $\pm$ 0.02\\
\hline
log SFR$_{\rm AGNs}$ & $-$2.41 $\pm$ 0.02 & $-$2.08 $\pm$ 0.01 &  $-$2.12 $\pm$ 0.02 & $-$1.96 $\pm$ 0.01 & $-$2.47 $\pm$ 0.03 & $-$2.48 $\pm$ 0.02 & $-$2.70 $\pm$ 0.04 & $-$3.00 $\pm$ 0.02\\
\hline
\hline
$D_{\rm n}4000$$_{\rm SFGs}$ & 1.22 $\pm$ 0.01 & 1.34 $\pm$ 0.005 & 1.34 $\pm$ 0.01 & 1.32 $\pm$ 0.014 & 1.27 $\pm$ 0.015 & 1.11 $\pm$ 0.018 & 1.12 $\pm$ 0.020 & 1.13 $\pm$ 0.003\\
\hline
log SFR$_{\rm SFGs}$ & $-$1.92 $\pm$ 0.04 & $-$2.04 $\pm$ 0.01 &  $-$2.03 $\pm$ 0.02 & $-$2.04 $\pm$ 0.03 & $-$1.90 $\pm$ 0.01 & $-$1.86 $\pm$ 0.02 & $-$1.82 $\pm$ 0.01 & $-$1.75 $\pm$ 0.03\\
\hline
\end{tabular}
Notes: In this table, we present the average values of $D_{\rm n}$4000/SFR for the two samples, as well as the average values for each ring. The errors show the standard deviation of the mean.
\end{center}
\end{table}

\section{Visual Overlap Between the $D_{\rm n}$4000 and the H$\alpha$ Arc Structure}\label{visual overlap}
We combined $D_{\rm n}$4000 with H$\alpha$ and found that the position where $D_{\rm n}$4000 decreased exactly corresponded to the peak of H$\alpha$ emission. In Fig. \ref{Arc_check}, we present the overlap of the H$\alpha$  and DECaLS r-band optical images.

\begin{figure*}[!ht]
 \centering
 \subfigure[]
 {
  \begin{minipage}{0.31\textwidth}
   \centering
   \includegraphics[width=\linewidth]{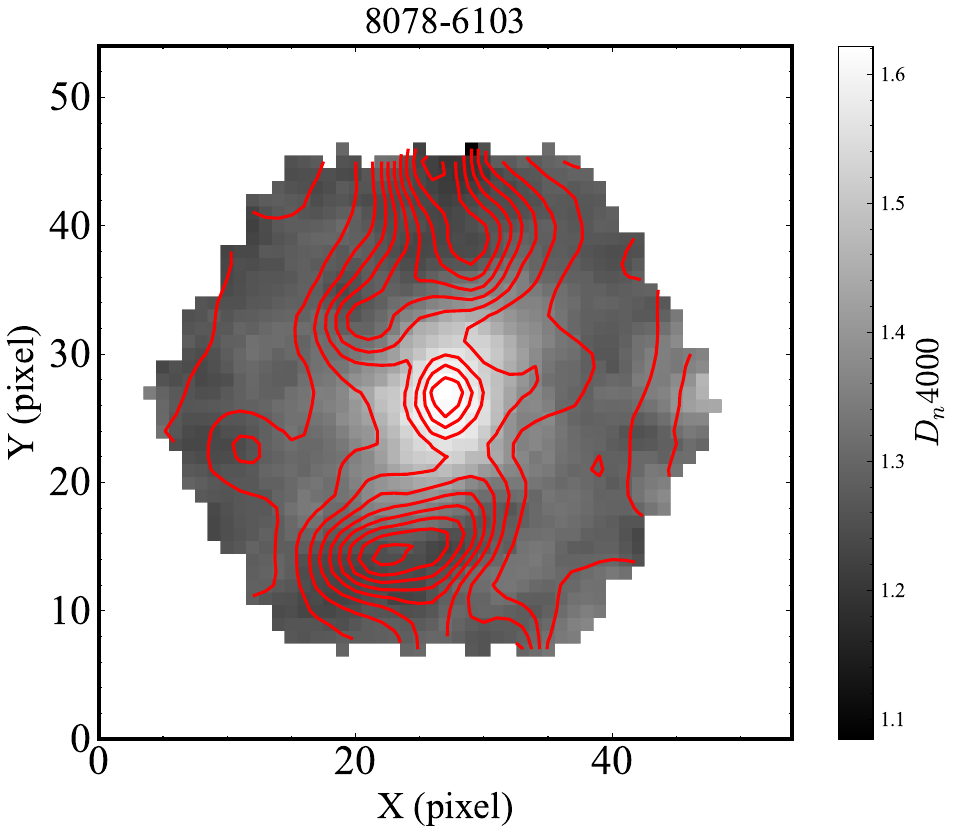}
   \label{App_B_f1}
  \end{minipage}
 }
    \subfigure[]
    {
     \begin{minipage}{0.31\textwidth}
      \centering
      \includegraphics[width=\linewidth]{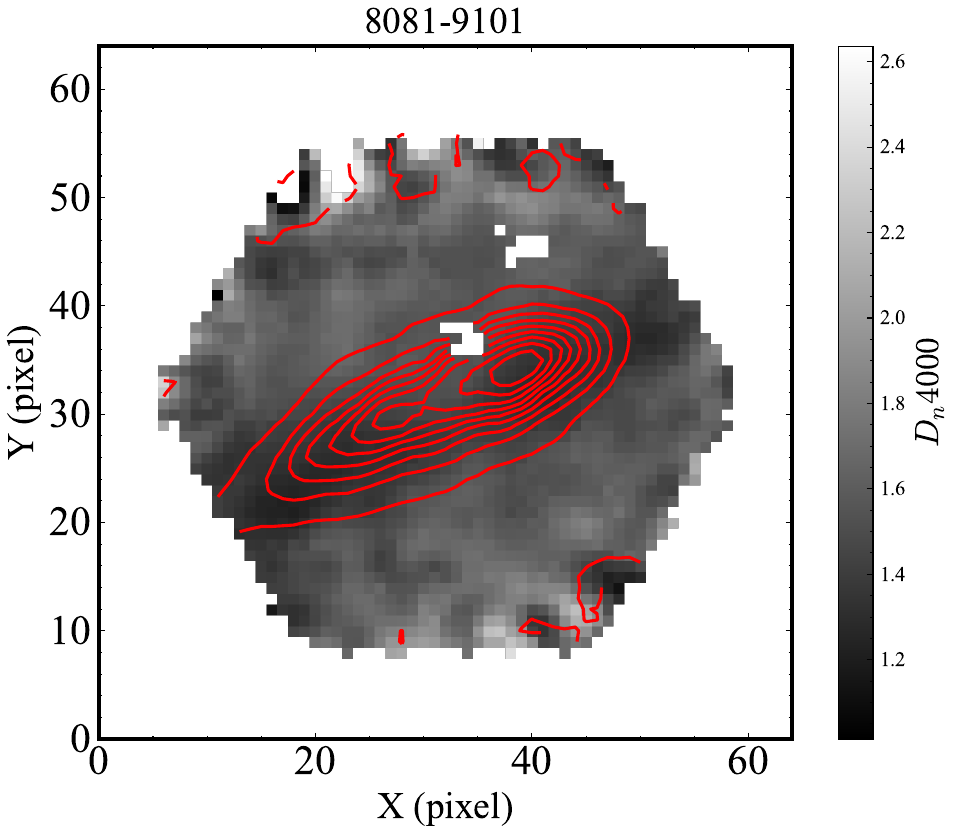}
      \label{App_B_f2}
     \end{minipage}
    }
    \subfigure[]
    {
     \begin{minipage}{0.31\textwidth}
      \centering
      \includegraphics[width=\linewidth]{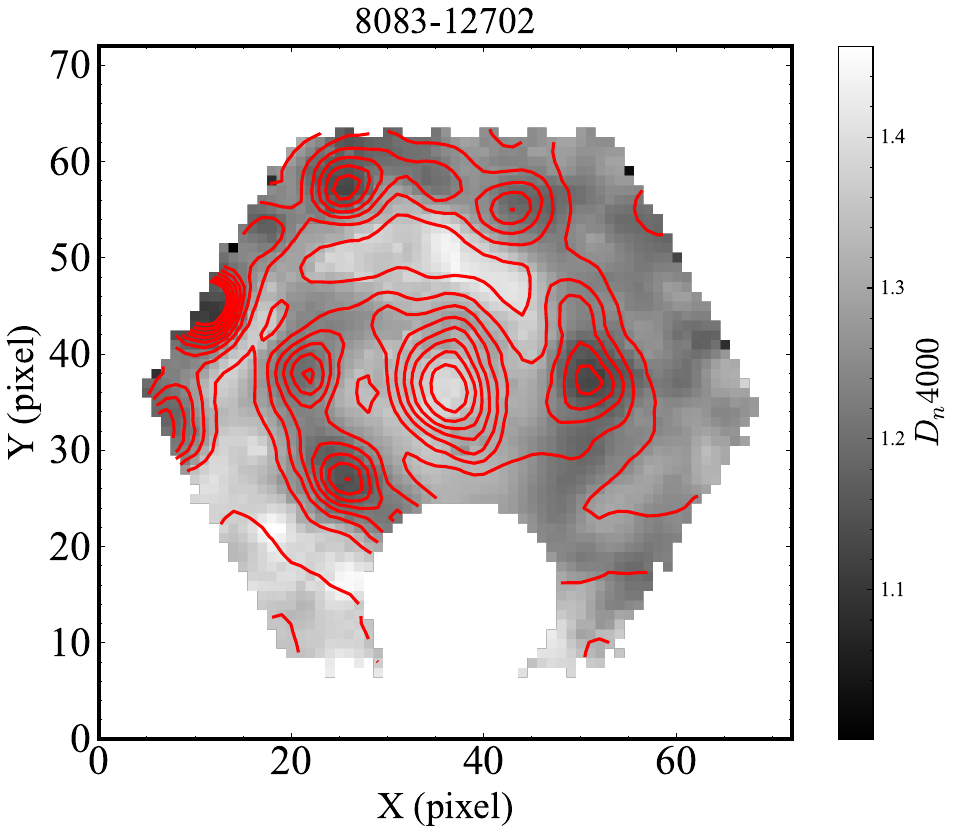}
      \label{App_B_f3}
     \end{minipage}
    }
     \subfigure[]
    {
     \begin{minipage}{0.31\textwidth}
      \centering
      \includegraphics[width=\linewidth]{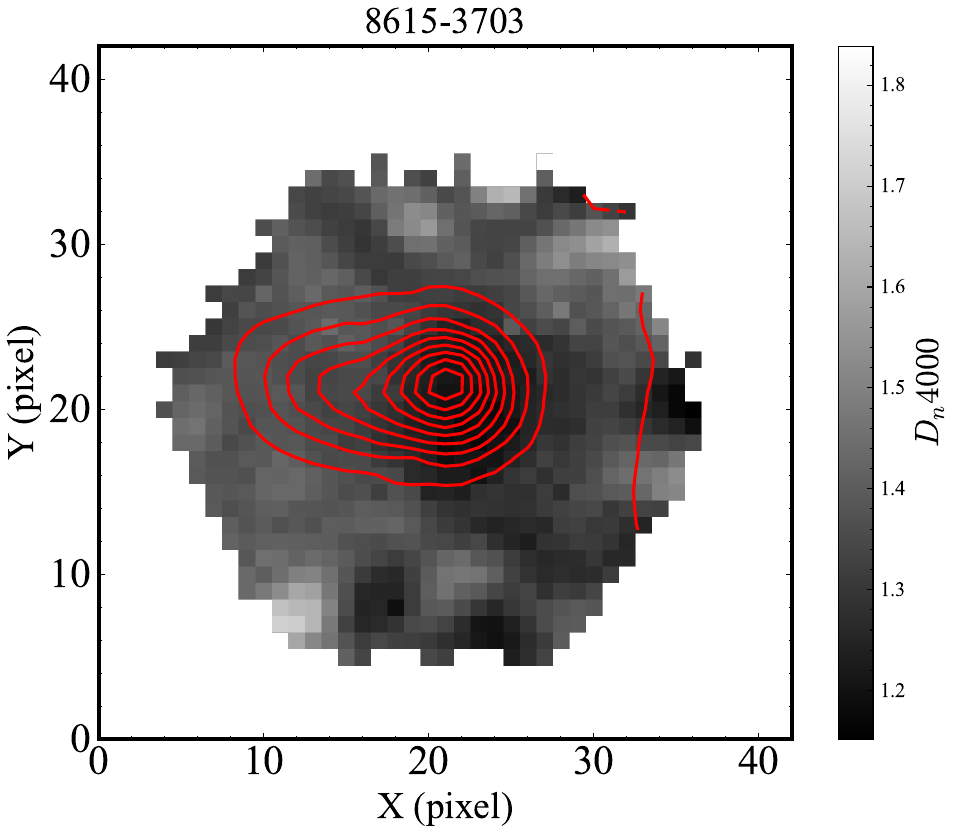}
      \label{App_B_f4}
     \end{minipage}
    }
     \subfigure[]
    {
     \begin{minipage}{0.31\textwidth}
      \centering
      \includegraphics[width=\linewidth]{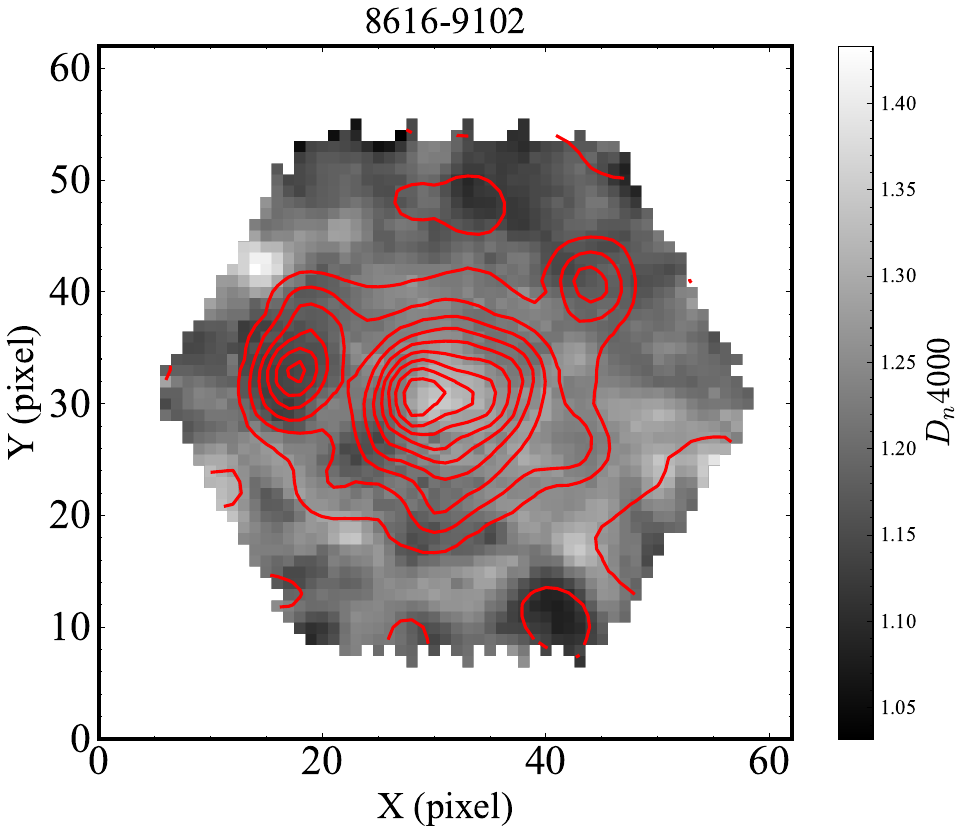}
      \label{App_B_f5}
     \end{minipage}
    }
     \subfigure[]
    {
     \begin{minipage}{0.31\textwidth}
      \centering
      \includegraphics[width=\linewidth]{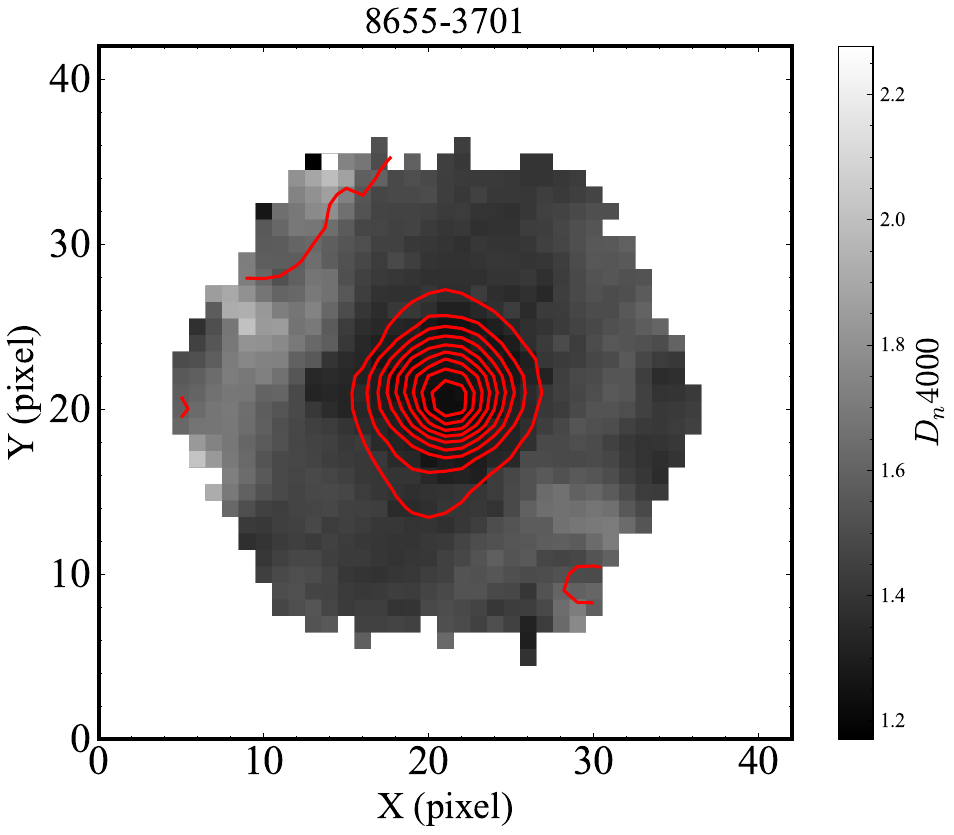}
      \label{App_B_f6}
     \end{minipage}
    }
\caption{ The relative positions of the H$\alpha$ contours (red lines) and $D_{\rm n}$4000 (gray) for our AGN hosts.}
\label{Arc_Sps}
\end{figure*}

\begin{figure*}[!ht]
 \centering
 \subfigure[]
 {
  \begin{minipage}{0.31\textwidth}
   \centering
   \includegraphics[width=\linewidth]{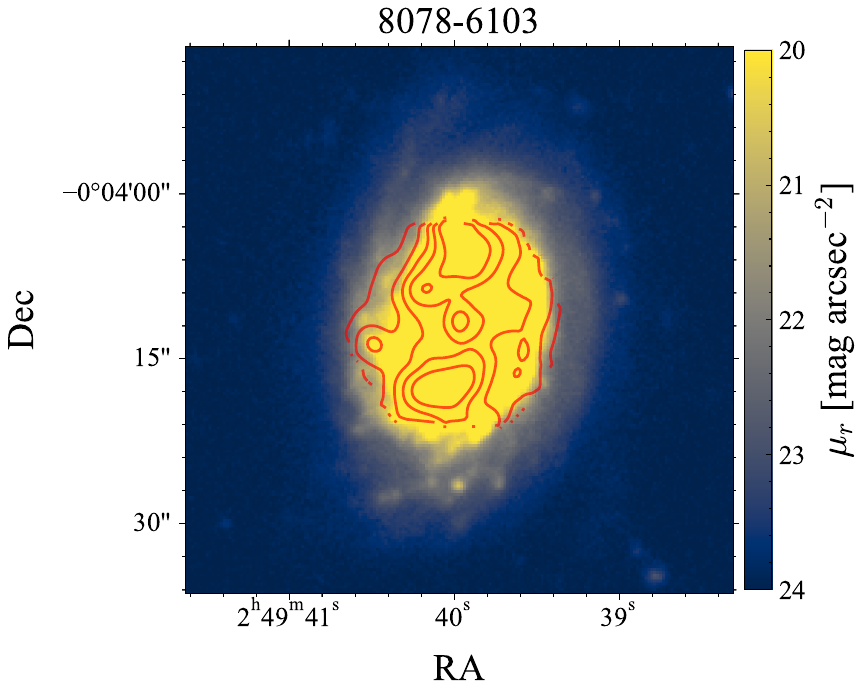}
   \label{App_C_f1}
  \end{minipage}
 }
    \subfigure[]
    {
     \begin{minipage}{0.31\textwidth}
      \centering
      \includegraphics[width=\linewidth]{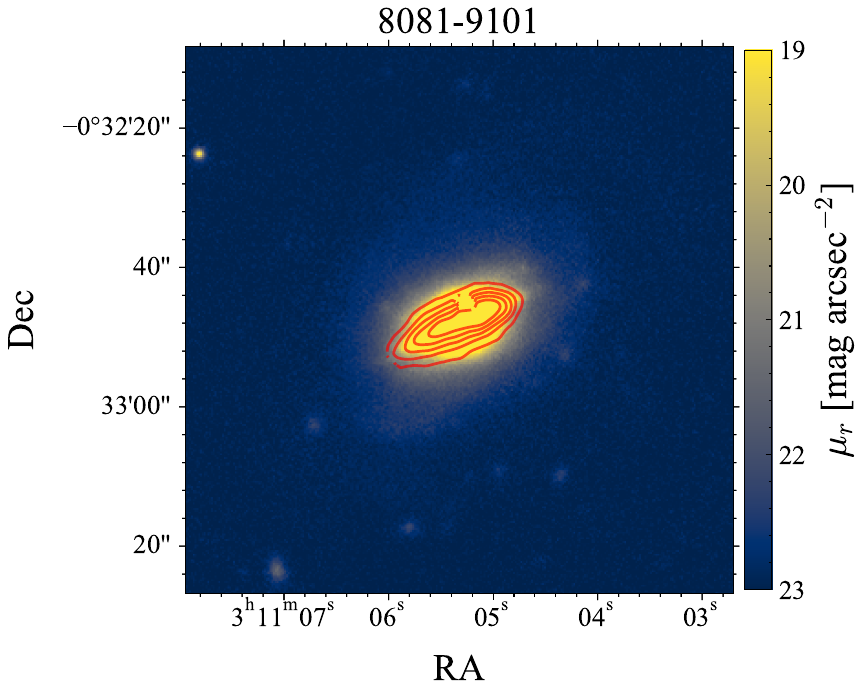}
      \label{App_C_f2}
     \end{minipage}
    }
    \subfigure[]
    {
     \begin{minipage}{0.31\textwidth}
      \centering
      \includegraphics[width=\linewidth]{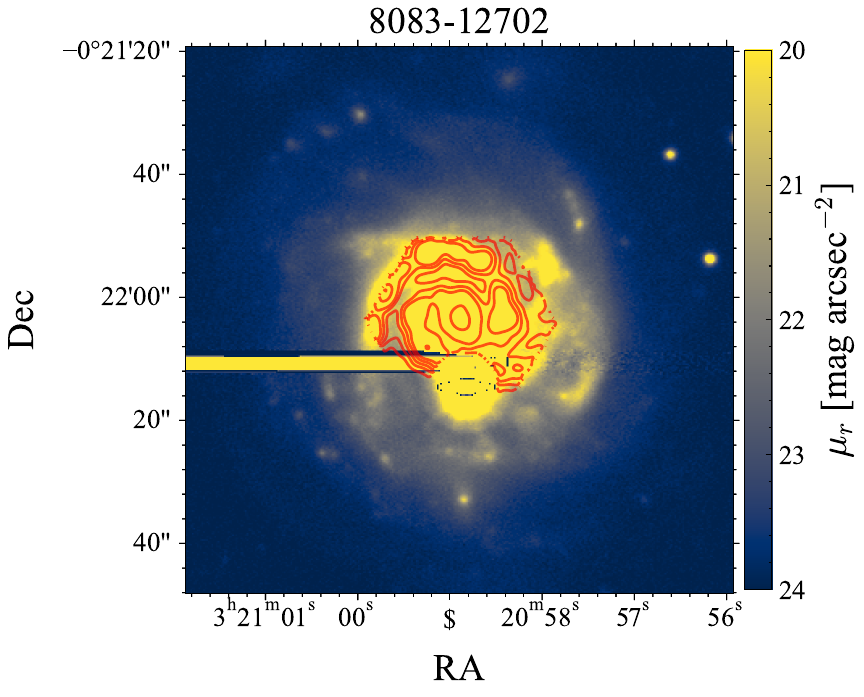}
      \label{App_C_f3}
     \end{minipage}
    }
     \subfigure[]
    {
     \begin{minipage}{0.31\textwidth}
      \centering
      \includegraphics[width=\linewidth]{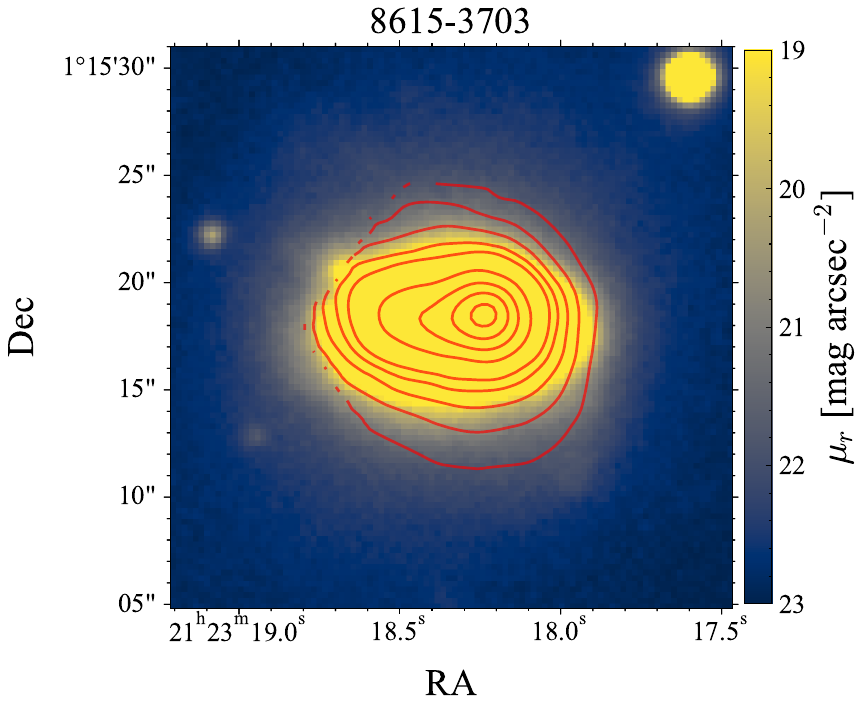}
      \label{App_C_f4}
     \end{minipage}
    }
     \subfigure[]
    {
     \begin{minipage}{0.31\textwidth}
      \centering
      \includegraphics[width=\linewidth]{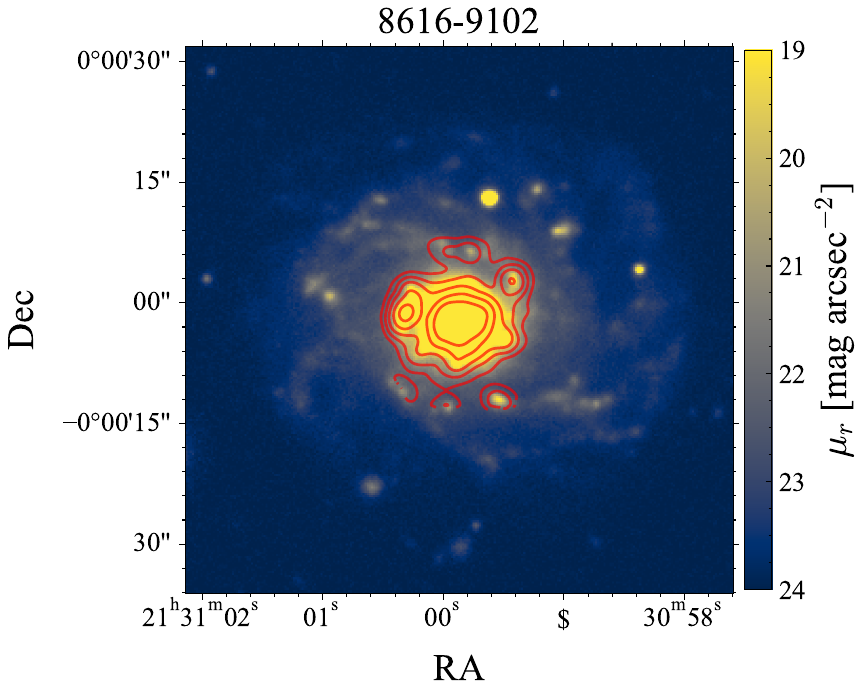}
      \label{App_C_f5}
     \end{minipage}
    }
     \subfigure[]
    {
     \begin{minipage}{0.31\textwidth}
      \centering
      \includegraphics[width=\linewidth]{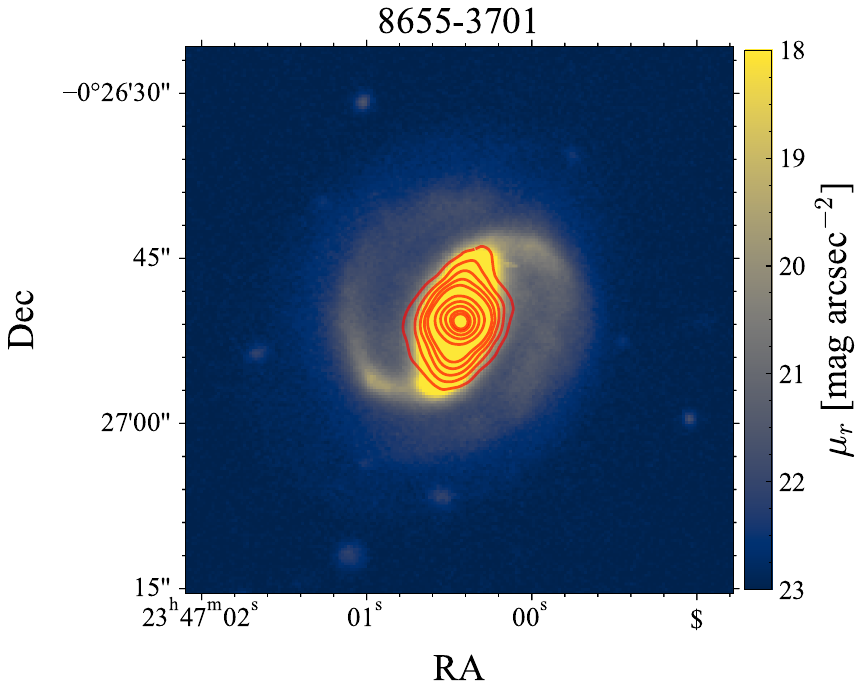}
      \label{App_C_f6}
     \end{minipage}
    }
\caption{The visual overlap between H$\alpha$ emissions (red contours) between DECaLS r-band images (yellow). We found that the peak of H$\alpha$ exactly corresponds to the spiral arms of the galaxy, especially in cases of 8078-6103, 8083-12702, and 8616-9102. Of course, for galaxy 8655-3701, it could also be caused by the bar structure within the galaxy.}
\label{Arc_check}
\end{figure*}

\end{appendix}

\end{CJK}
\end{document}